\documentclass[11pt]{article}
\usepackage{amsmath,amssymb,amsfonts,graphicx,slashed,color,mathtools,amsthm}
\usepackage{jheppub}
\usepackage{comment}
\usepackage{enumerate}
\usepackage{float}
\usepackage{caption}
\usepackage{subcaption}
\usepackage{pdflscape}
\usepackage{appendix}

\newcommand{\be}{\begin{equation}}
\newcommand{\ee}{\end{equation}}
\newcommand{\bea}{\begin{eqnarray}}
\newcommand{\eea}{\end{eqnarray}}
\newcommand{\beas}{\begin{eqnarray*}}
\newcommand{\eeas}{\end{eqnarray*}}
\newcommand{\ba}{\begin{array}}
\newcommand{\ea}{\end{array}}

\newcommand{\ket}[1]{\left| #1 \right\rangle}
\newcommand{\abs}[1]{\left| #1 \right|}
\newcommand{\bra}[1]{\left\langle #1 \right|}
\newcommand{\avg}[1]{\left\langle #1 \right\rangle}

\title{Ordinary wormholes}

\author[1,2,3]{Alexander Maloney}
\author[1]{Viraj Meruliya}
\author[4]{Mark Van Raamsdonk}
\affiliation[1]{Department of Physics, McGill University,
        Montr\'eal, QC H3A 2T8, Canada}
\affiliation[2]{Department of Physics, Syracuse University Syracuse, NY, 13244, USA}
\affiliation[3]{Institute for Quantum and Information Sciences, Syracuse University, Syracuse, NY, 13244, USA}
\affiliation[4]{Department of Physics and Astronomy, University of British Columbia,
6224 Agricultural Road, Vancouver, B.C.\ V6T 1Z1, Canada}

\emailAdd{mav@phas.ubc.ca}
\emailAdd{viraj.meruliya@mail.mcgill.ca}
\emailAdd{admalone@syr.edu}

\abstract{Euclidean wormholes have played a key role in the recent ``disorder averaged" approaches to quantum gravity and holography, but are typically only considered in somewhat special theories of gravity, such as theories in low dimensions or theories with exotic matter content (such as axions).  These exotic theories have advantage that both the matter and gravitational sectors can be treated completely classically.  However, once this constraint is relaxed we find that Euclidean wormholes arise generically, with no special constraints on the matter content. The key point is that there is a self-consistent approximation where the metric is treated classically but matter is treated quantum mechanically.  The resulting wormholes are {\it ordinary} in the sense that they rely on the usual approximations used in, for example, the construction of star or FRW solutions in general relativity.  Indeed, these are the Euclidean continuations of ordinary FRW solutions with big bang/crunch singularities. We describe several examples of these ordinary wormholes and discuss the relation to existing constructions and the holographic interpretation in terms of a dual CFT.}

\begin{document}

\maketitle

\section{Introduction}

In this paper, we explore the construction and interpretation of a class of Euclidean wormhole spacetimes. We have two separate motivations for this.

First, recent work (see for example \cite{Saad:2019lba, Almheiri:2019qdq,Penington:2019kki,Marolf:2020xie,Belin:2020hea,Maloney:2020nni,Chandra_2022,Chen:2020tes,VanRaamsdonk:2020tlr,Marolf:2021kjc,Saad:2021rcu,Cotler:2020ugk, Cotler:2021cqa,Balasubramanian:2022gmo, Pollack:2020gfa}) has highlighted that the presence of Euclidean wormhole solutions provides important insights into the nature of semiclassical gravity and its relation to microscopic descriptions.  Despite much effort, attempts to construct fully UV complete theories of quantum gravity based on Euclidean path integrals seems to work in only a few very simple settings. 
Wormholes have emerged as a ``smoking gun" signature of the failure of the gravitational path integral to define a fully consistent UV theory. This is particularly clear in the context of the AdS/CFT correspondence, where the existence of wormholes with multiple boundaries leads to apparent non-factorization of the corresponding dual CFT. It is believed that factorization can be restored only once the correct non-perturbative bulk degrees of freedom are accounted for. More generally, the existence of wormhole solutions can significantly modify conclusions arising from perturbative gravity calculations in positive as well as negative ways. For example, in the black hole information puzzle wormholes are the key contributions which demonstrate that the semi-classical gravitational path integral is consistent with a finite number of black hole microstates \cite{Saad:2019lba,Almheiri:2019qdq,Penington:2019kki,Maloney:2020nni,Balasubramanian:2022gmo,Balasubramanian:2022lnw, Iliesiu:2024cnh, Boruch:2024kvv}. In holographic setups, wormholes can indicate a natural statistical variance in certain observables; this may indicate that the underlying microscopic description involves an ensemble of theories \cite{Cotler:2016fpe,Saad:2018bqo,Saad:2019lba, Afkhami-Jeddi:2020ezh, Maloney:2020nni, Chandra_2022, Chandra:2024vhm}, or reveal natural statistical ensembles (e.g. of operators with dimensions near a certain value) present within individual theories \cite{Pollack:2020gfa,Belin:2020hea,Cotler:2021cqa,Altland:2021rqn,Chandra:2022fwi}. 

A second motivation for our study is the desire to model cosmological spacetimes using holography. When $\Lambda  < 0$, the gravitational effective field theories associated with holographic CFTs have solutions describing  (approximately) homogeneous and isotropic big bang cosmologies (see \cite{VanRaamsdonk:2024sdp} for a recent discussion).  A central challenge is to give a CFT interpretation for these cosmologies.
These cosmological solutions typically have an analytic continuation to Euclidean time which is a wormhole with two asymptotically AdS boundaries, suggesting a holographic construction in terms of a pair of Euclidean CFTs. So one route towards the construction of a holographic theory of cosmology is to come up with a specific CFT construction that describes such a wormhole, and to then define cosmological observables via analytic continuation \cite{Maldacena:2004rf, Antonini:2022ptt, Antonini:2022fna, Sahu:2024ccg}.\footnote{Equivalently, we can think of the Euclidean path integral as constructing a state for the Lorentzian cosmology.} A significant challenge here is to understand a detailed CFT construction that will lead to the wormhole, preferably as the dominant saddle.

The construction of asymptotically AdS Euclidean wormhole solutions is notoriously challenging: constructions often require exotic ingredients such as axion matter  or negative spatial curvature, and when solutions exist, there are often instabilities or lower-action saddles with the same boundary conditions that would dominate in the gravitational path integral \cite{Arkani-Hamed:2007cpn,Maldacena:2004rf,Betzios:2019rds, Marolf:2021kjc}. In contrast, $\Lambda < 0$ cosmological solutions appear to be very generic. Starting with any spatial curvature and any combination of matter and radiation density with sufficient energy density,\footnote{We require $\rho > d(d-1)/(16 \pi G)(K + 1)$ where $K$ is the spatial curvature.} we can find a big bang/big crunch solutions to the Friedman equation that analytically continues to a Euclidean wormhole, at least at the level of background cosmology. 

These observations motivate us to ask whether there exists a more generic class of Euclidean wormholes related by analytic continuation to some of these cosmological solutions.\footnote{A caveat is that time-reversal symmetry is required for the analytic continuation to give a real Euclidean wormhole solution, while cosmologies are typically not time-reversal symmetric once perturbations about the background are considered. The solutions we consider in this paper will also have time-symmetric perturbations.} Compared to the wormhole constructions typically considered, the key difference is that we will explicitly include matter and/or radiation rather than looking for a solution supported by homogeneous and isotropic classical background fields. 

The main point of this paper is to present a broad class of ``inhomogeneous'' wormhole solutions of this type and to discuss their interpretation in terms of CFT physics.

\subsubsection*{Wormholes from heavy particle cosmologies}

In Section \ref{sec:massive}, we consider solutions where the matter is a collection of heavy particles. In order to find analytic solutions, we work in 2+1 dimensions where a massive particle corresponds to a conical defect in the spacetime. Via a gluing procedure, we describe a very general class of solutions with arbitrary spatial topology and very general distributions of matter particles, generalizing wormholes considered recently in \cite{Chandra_2022}.

For all of the solutions, the Lorentzian spacetime geometry is simply
\begin{equation}
\label{Lor1}
    ds^2 = -dt^2 + \cos^2(t/\ell) d \Sigma^2
\end{equation}
where $d \Sigma^2$ is locally hyperbolic space with defects. The corresponding Euclidean wormhole wormhole solutions are found by taking $t_E = i t$; these Euclidean wormholes have two boundaries at $t_E \to \pm \infty$. For closed universe examples, we find a universal formula for the total mass
\begin{equation}
\label{Mass1}
M_{tot} = {1 \over 4 G} ({A_{tot} \over 2 \pi \ell^2} + \chi)
\end{equation}
in terms of the area of the spatial slice at the time-symmetric point and the Euler characteristic $\chi$. This gives precisely the mass density predicted from the homogeneous and isotropic Friedmann equation based on the scale factor $a(t) = \cos(t/\ell)$, so the inhomogeneities have no effect on the scale factor evolution.

The classical solutions we consider approximate more microscopic semiclassical configurations where the particles are described explicitly by quantum fields. In Section \ref{sec:QFT}, we discuss the explicit description of the solutions away from the heavy particle limit where this quantum field theory description is required. Here, a key point is that there is {\it not} a good description of the solutions in terms of classical fields. In the Lorentzian picture, the quantum field theory states sourcing the cosmology have a significant variance for the matter fields relative to their expectation values (which are simply zero in some case). In the Euclidean picture, there is not a single classical configuration that serves as a dominant saddle, but rather many nearby configurations related by small variations in the matter fields. Nevertheless, the variance of the stress-energy tensor for these quantum fields can be small, allowing a metric field with small variance that can be well-approximated by a classical solution. Thus, we find it plausible that the class of wormhole solutions that we have described explicitly in the heavy particle limit generalizes to arbitrary particle masses, but the description of such solutions necessarily required a quantum treatment of the fields.

\subsubsection*{CFT Interpretation}

In Section \ref{sec:CFT}, we consider the CFT interpretation of the massive particle wormholes and the associated cosmologies. The boundary geometry for the wormholes, with two disconnected components and a set of heavy particle worldlines intersecting these boundaries, suggests that the CFT construction involves a CFT on living on the disconnected boundary geometry with a set of operators inserted at the locations corresponding to the worldline intersections.

A precise relation between this naively factorized correlator and the connected wormhole geometry for simple cases of this construction was described in \cite{Chandra_2022}.  In that work, it was shown that the gravitational action for the wormhole solution corresponds to a ``coarse-grained'' CFT calculation in which products of OPE coefficients appearing in the calculation are replaced by their average in a Gaussian ensemble of such OPE coefficients. 

\begin{figure}
 \centering
    \includegraphics[scale = 0.3]{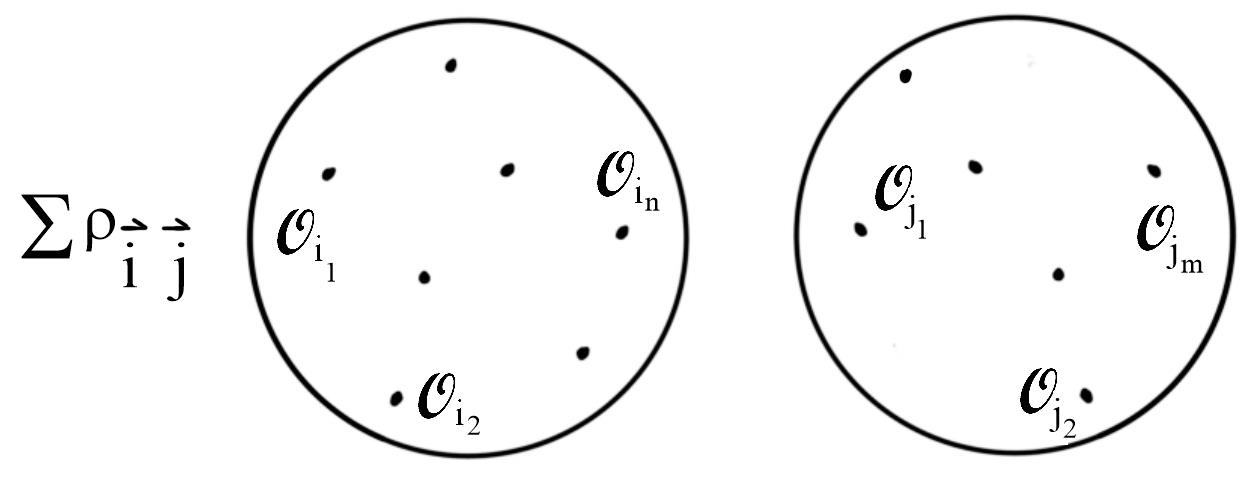} 
    \caption{Disconnected CFT correlator with an ensemble of operator insertions. The connected part (defining the variance of an ensemble of complicated single-CFT correlators) is calculated by a wormhole contribution in the dual gravity theory that analytically continues to a cosmological solution.}
    \label{fig:TFD3intro}
\end{figure}

Our construction is reflection-symmetric, so the wormholes are naturally associated with the connected part of the ensemble average of the product of two copies of a CFT correlator with many operator insertions. This defines the variance in the ensemble of a single CFT correlator with many insertions.  Here, the ensemble might be over possible CFT data, but since the correlator itself has a large amount of data - the type and location of all the operator insertions, in some cases we can define a natural ensemble of such correlators within a single CFT, where the ensemble is over the details of the operator insertions (e.g. the precise operator that creates a given heavy particle) - see Figure \ref{fig:TFD3intro}.

The existence of the massive particle wormhole solutions that we construct in this paper suggests that all of these considerations apply not only to observables involving heavy CFT operators, but also to complicated CFT observables with a large number of operators that are not necessarily heavy or even close to the black hole threshold. By extrapolation, it seems plausible that wormhole contributions (and the universal variance of CFT correlators that they give rise to) appear even for correlators involving light operators provided that the total energy of the particles they produce is of order $c$.

This can be seen as a simple generalization of the results in \cite{Chandra_2022}, where it was shown that a simple wormhole with three particle wordlines led to a non-trivial variance for the three point function 
$\overline{\langle {\cal O}_1 {\cal O}_2 {\cal O}_3\rangle^2}$ when the operators ${\cal O}_i$ all have dimension of order $c$.
We now see that if we take a collection of  $n$ operators ${\cal O}_i$ which are light, in the sense that their dimensions are of order $1$, there will still be a variance for the $n$-point function 
$\overline{\langle {\cal O}_1 \dots {\cal O}_n\rangle^2}$ as long as $n$ is of order $c$.  In other words, even the lightest operators will be subject to statistical averaging when coupled to gravity, but this averaging is only evident when considering highly refined observables, such as $n$-point functions when $n\approx \frac{\ell}{G}$.\footnote{This is somewhat reminiscent of the fact that even in weakly coupled theories, sufficiently complicated observables can still be effectively strongly coupled.  For example, in QED $2\to2$ scattering amplitudes can be treated using weak-coupling expansions due to the smallness of the fine structure constant $\alpha \sim \frac{1}{137}$.  But nevertheless scattering processes involving ${\cal O}(137)$ electrons do not admit a similar weak-coupling expansion.}

\subsubsection*{Sinusoidal scalar cosmologies}

In section \ref{sec:radiation}, we investigate a different type of wormhole/cosmology solution, originally discussed by Marolf and Santos in \cite{Marolf:2021kjc} for the wormhole case. Here, we have radiation described by scalar fields with sinusoidal spatial variation. A particular combination of three complex scalar fields with sinusoidal spatial variation gives a stress-energy tensor that is homogeneous and isotropic, so it is consistent to take the exact metric to be of FRW form. In the Lorentzian picture, we find big-bang/big crunch cosmological solutions for arbitrary wavelengths. Interestingly, not all of these have analytic continuations to asymptotically AdS wormholes. When the scalar wavelengths at the time-reflection symmetric slice are below a critical value of order the AdS scale, the Euclidean solution develops singularities at times $\tau = \pm \tau_s$ rather than asymptotically AdS regions at $\tau = \pm \infty$. This example illustrates that certain cosmological solutions that would naively continue to wormholes actually give singular Euclidean solutions.

\section{Setup and Generalities}
\label{sec:setup}

In this paper, we consider asymptotically AdS wormhole solutions and the associated cosmological solutions obtained from these by analytic continuation. The Euclidean solutions are described by a metric
\begin{equation}
    ds^2 = d \tau^2 + a_E(\tau)^2 d \Sigma^2
\end{equation}
where $a_E(\tau) \sim e^{|\tau|/\ell}$ for $\tau \to \pm \infty$. We will focus on examples where the geometry has a time-reflection symmetry $\tau \to - \tau$ with a minimum scale factor at the reflection-symmetric point $\tau = 0$. We can choose $a_E(\tau = 0)=1$ without loss of generality.

The Lorentzian solutions are obtained by analytic continuation of the time direction. These have the standard FRW form
\begin{equation}
    ds^2 = -dt^2 + a(t)^2 d \Sigma^2
\end{equation}
where $a(i \tau) = a_E (\tau)$. Since we are considering gravitational effective theories with negative cosmological constant, these associated cosmologies are big bang / big crunch cosmologies where $a(t) \to 0$ for $t \to \pm t_0$. In our conventions, $t=0$ represents recollapse point with maximal scale factor $a=1$. 

In studying cosmological solutions, it is typical to model the stress-energy tensor for matter/radiation as a homogeneous and isotropic perfect fluid with energy density $\rho$ and pressure $p$. Then, the evolution of the scale factor is governed by the Friedmann equation
\begin{equation}
    \left({\dot{a} \over a}\right)^2 + {K \over a^2} = {16 \pi G \over d(d-1)} \rho - {1 \over \ell^2} \; .
\end{equation}
where $K$ is the Gaussian curvature of the spatial geometry, which is assumed to be homogeneous and isotropic.


Demanding that the Friedmann equation is satisfied at the recollapse point in the cosmology gives
\begin{equation}
   \rho = {d(d-1) \over 16 \pi G} ( K + {1 \over \ell^2})  
\end{equation}
so the energy density is fixed in terms of the spatial curvature. 

We see that apart from the special case $K= -1/\ell^2$ that corresponds to pure AdS, the existence of the time-symmetric cosmological background requires a nonvanishing density of matter and/or radiation (and $K \ge -1/\ell^2$). This matter/radiation must also be present in the associated Euclidean wormhole solution. 

Within the perfect fluid approximation, constructing cosmology/wormholes solutions is very easy. For any $\rho(a)$, we can write the associated Lorentzian scale factor implicitly as
\begin{equation}
    |t(a)| = \int_a^1 {d \hat{a} \over \hat{a} \sqrt{{16 \pi G \over d(d-1)} \rho(\hat{a}) - {1 \over \ell^2} - {K \over \hat{a}^2}}}
\end{equation}
while the Euclidean scale factor for the associated cosmology is
\begin{equation}
    |\tau(a_E)| = \int_1^a {d \hat{a} \over \hat{a} \sqrt{-{16 \pi G \over d(d-1)} \rho(\hat{a}) + {1 \over \ell^2} + {K \over \hat{a}^2}}}
\end{equation}

In this paper, we would like to consider solutions that go beyond the perfect fluid approximation, where the matter is represented explicitly via particles or fields.

In the next sections, we consider solutions where the energy density arises from massive particles (Section \ref{sec:massive}), from black holes (Section \ref{sec:heavy}), and from classical radiation (Section \ref{sec:radiation}).

\section{Massive particle wormholes cosmologies}
\label{sec:massive}

In this section, we construct exact solutions for Euclidean wormholes and associated cosmologies in 2+1 dimensions where the matter consists of a collection of particles. The solutions are locally AdS${}_3$ away from the particles, which give rise to conical defects in the geometry with deficit angle
\begin{equation}
\label{DeltaM}
\delta = 8 \pi G M 
\end{equation}
where $M$ is the particle mass.  In previous discussions of such wormholes (such as in \cite{Chandra_2022}), the defecit angles have typically been taken to have $\delta={\cal O}(1)$; in the dual CFT interpretation, these particles are dual to operators with dimension of order $c$.  We will allow the deficit angles much smaller than one, $\delta \ll 1$.  However, as we will discuss below, we must still require that $M$ is large in AdS units so that the particles can be well localized in AdS and therefore described by point particles.  In the dual CFT language this means that the dimensions are much larger than ${\cal O}(1)$, but can be much smaller than $c$.\footnote{More precisely, they will have dimensions of order $\epsilon c$, where $\epsilon$ is an arbtrarily small (but fixed) positive number in the large $c$-limit.}

\subsection{Spatial geometry}

The cosmological solutions and wormhole solutions will be time-reflection symmetric. We begin by describing the geometry of the spatial slice that is fixed under this reflection. This geometry is common to both solutions. By the time-reversal symmetry, we expect this slice is locally an extremal surface in the $AdS_3$, so the local geometry is two-dimensional hyperbolic space away from the defects. The full geometry can be constructed by gluing together hyperbolic triangles, as we now describe.

\subsubsection*{Hyperbolic triangles}

Starting from two-dimensional hyperbolic space, we can construct triangles whose sides are geodesics. There is a unique triangle for each set of side lengths $(a,b,c)$ satisfying the usual triangle inequalities.

\begin{figure}
 \centering
    \includegraphics[scale = 0.5]{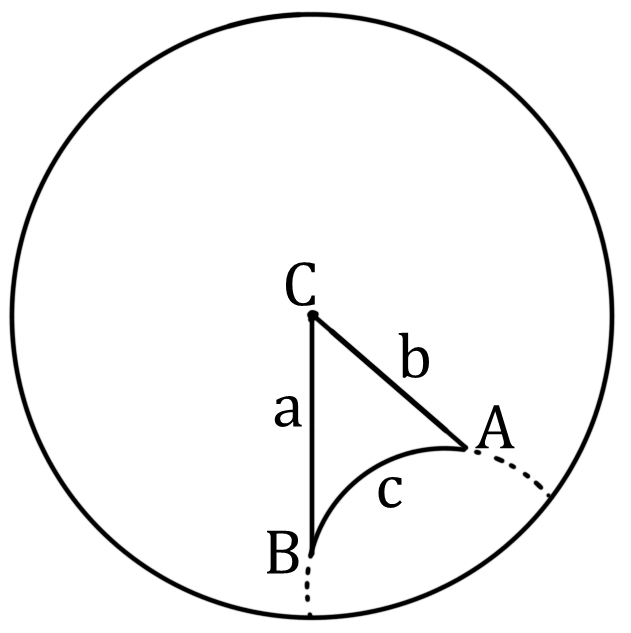}
    \caption{A geodesic hyperbolic triangle in the Poincar\'e disk}
    \label{fig:hyptriangle}
\end{figure} 

Alternatively, there is a unique triangle for each set of angles $(A,B,C)$ with $A+B+C < \pi$. The angles and side lengths can be related by the hyperbolic cosine law
\begin{equation}
\label{hypcos}
    \cos(C) = {\cosh(a/\ell) \cosh(b/\ell) - \cosh(c/\ell) \over \sinh(a/\ell) \sinh(b/\ell)} \;.
\end{equation}
where $-1/\ell^2$ is the Gaussian curvature of the hyperbolic space ($\ell$ will be the AdS length).

The hyperbolic triangles can be constructed starting with the Poincar\'e disk with metric
\begin{equation}
    ds^2 = { 4 \ell^2 (dx^2 + dy^2) \over (1 - x^2 - y^2)^2} \; .
\end{equation}
Geodesics are straight lines or circles in the $(x,y)$ plane that intersect the boundary perpendicularly. We can choose one vertex of the triangle at the center of the circle so that two sides of the triangle are segments on the radial geodesics. To construct the triangle with side lengths $(a,b,c)$, we can choose the angle between these straight line geodesics via (\ref{hypcos}), and the lengths of these segments to be $a$ and $b$. The remaining side lies on the unique circle that passes through the vertices $A$ and $B$ and is perpendicular to the boundary of the disk. This construction is depicted in Figure (\ref{fig:hyptriangle}).

\subsubsection*{Gluing triangles}

\begin{figure}
 \centering
    \includegraphics[scale = 0.5]{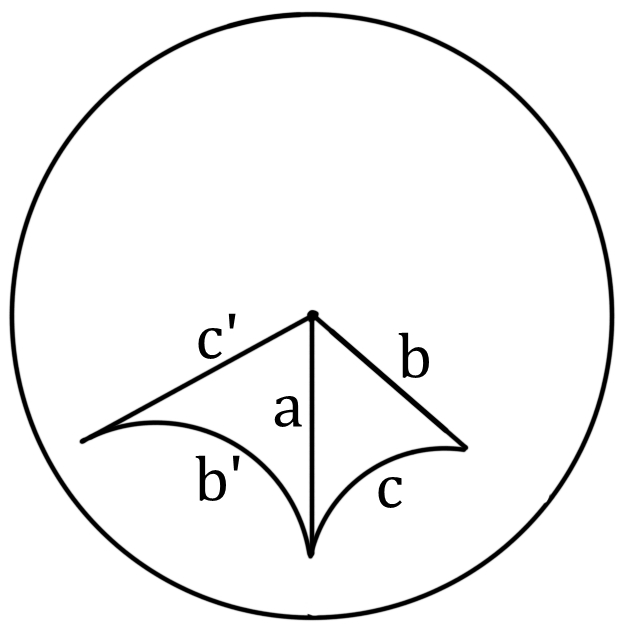}
    \caption{Gluing hyperbolic triangles.}
    \label{fig:hyp2}
\end{figure}

We can construct a very general solution starting from a spatial geometry that is constructed by gluing together hyperbolic triangles.\footnote{We could also use other polygons, but these can be further decomposed into triangles.} Any two hyperbolic triangles that have one side length in common can be glued smoothly together along equal length sides, for example using the construction shown in figure (\ref{fig:hyp2}).

Now, consider any collection of hyperbolic triangles with sides $(a_i,b_i,c_i)$ and any rule for gluing the sides (including gluing two equal sides of the same triangle) such that all sides are glued to some other side. The resulting surface will be locally smooth and hyperbolic away from the vertices. We impose the additional constraint that the gluing gives an orientable surface.\footnote{It may also be interesting to consider non-orientable geometries; these will lead to single-sided Euclidean AdS geometries.} 

\begin{figure}
 \centering
    \includegraphics[scale = 0.5]{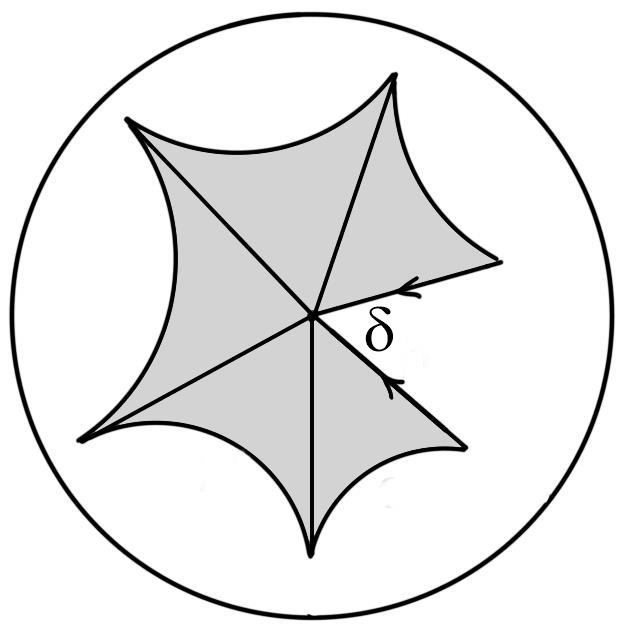}
    \caption{A vertex in the spatial slice, with deficit angle $\delta$.}
    \label{fig:hyp3}
\end{figure}

At each vertex, we require that the sum of the angles is less that or equal to $2 \pi$. Thus, we can always display the triangles glued at a vertex on a single copy of the Poincar\'e disk as shown in figure (\ref{fig:hyp3}). For vertices with a conical deficit, Einstein's equations demand that there is a particle with mass related to the deficit angle $\delta$ by (\ref{DeltaM}). 

For certain choices of the  side lengths $\{(a_i,b_i,c_i)\}$ and gluing there may be conical excesses at some vertices. In these cases, we can always scale up all the side lengths via $\{(\lambda a_i, \lambda b_i, \lambda c_i)\}$. The deficit angles all increase monotonically to $2\pi$ as $\lambda \to \infty$.\footnote{This follows because for any individual angle $C$ in a triangle with side lengths $(a,b,c)$, the cosine law gives $\lim_{\lambda \to \infty} C(\lambda a,\lambda b,\lambda c) =1$ and 
\begin{equation} 
    {d \over d \lambda} \cos(C(\lambda a,\lambda b,\lambda c)|_{\lambda = 1} = {\tanh(c/2) \sinh(c) \over \sinh(a) \sinh(b)}\left[{a \over \tanh(a)} + {b \over \tanh(b)} - {c \over \tanh(c/2)}\right] \; ,
\end{equation}
where we have set $\ell=1$. We can show that the expression in square brackets is positive by noting that the function $f(x) = x/\tanh(x)$ is convex, so 
\begin{equation}
    {a \over \tanh(a)} + {b \over \tanh(b)} = f(a) + f(b) \ge 2 f((a+b)/2) \ge 2 f(c/2) = {c \over \tanh(c/2)} \; .
\end{equation}
The last inequality follows from the triangle inequality on the side lengths.} Thus, for $\lambda$ greater than some $\lambda_0$ the geometry will have only non-negative deficit angles.

\subsection{Spacetime geometry}

\begin{figure}
 \centering
    \includegraphics[scale = 0.3]{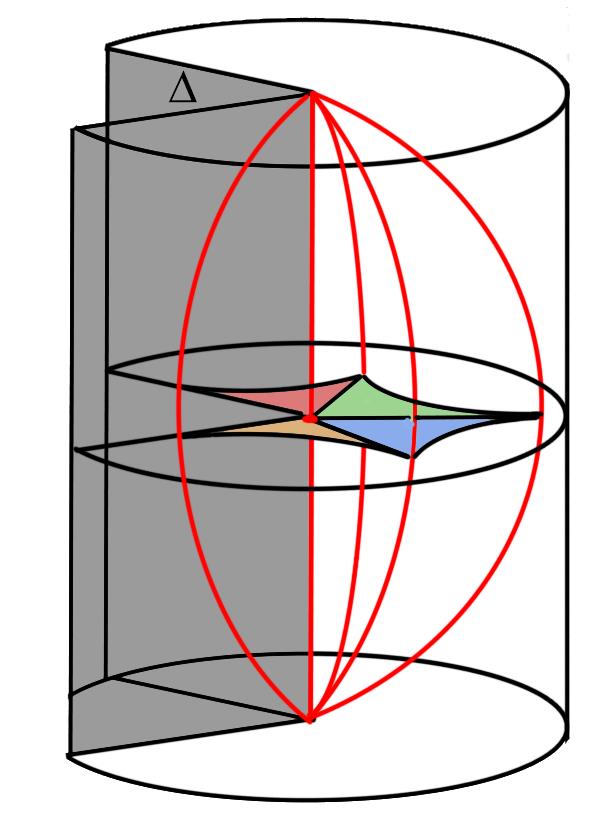}
    \caption{Spacetime regions in the Lorentzian cosmological solution associated with four hyperbolic triangles glued at a vertex. Red paths indicate the trajectories of the nearby massive particles. The convergence points correspond to the big bang and the big crunch.}
    \label{fig:ads_hyp}
\end{figure} 

By gluing hyperbolic triangles, we have defined a general class of locally hyperbolic spatial metrics $d \Sigma^2$. Given any one of these, we can obtain an exact solution Einstein's equations simply by taking
\begin{equation}
\label{Lor}
    ds^2 = -dt^2 + \cos^2(t/\ell) d \Sigma^2
\end{equation}
in the Lorentzian case and 
\begin{equation}
\label{Euc1}
    ds^2 = d\tau^2 + \cosh^2(\tau/\ell) d \Sigma^2
\end{equation}
in the Euclidean case. For $d \Sigma^2$ equal to the full two-dimensional hyperbolic space, these are Lorentzian and Euclidean $AdS_3$ in FRW coordinates. For our case with conical defects, these spacetimes are locally $AdS_3$ and thus satisfy the Einstein equations everywhere away from the defects. At the defects, the Einstein equations require that the appropriate relation (\ref{DeltaM}) between deficit angles and particle masses holds, but we choose each particle mass so that this is satisfied. The trajectories of these particles are consistent since they correspond to geodesics in the background AdS geometry.

A useful way to understand the spacetime geometry is by making use of the standard cylinder representation of global AdS as shown in figure (\ref{fig:ads_hyp}). With a massive particle at the center of AdS, we have a conical deficit, indicated by the missing wedge in the figure. The trajectories of the nearby massive particles correspond to geodesics with zero velocity at $t=0$. All such geodesics meet at the center of AdS after times $t = \ell {\pi \over 2}$. In the full cosmological spacetime, these points correspond to the big bang (if they meet in the past) and the big crunch (if they meet in the future).

\subsection{Matter density}

A remarkable feature of the solutions we have described is that in all cases, the matter density is precisely the same as for a homogeneous and isotropic solution with the same topology and scale factor evolution. This means that there are no backreaction effects associated with the inhomogeneities.

To show this, we recall that for any hyperbolic triangle, the area of the triangle is given simply in terms of the angles by the formula 
\begin{equation}
    Area/\ell^2 = \pi - \sum_{triangle } \theta_i \; .
\end{equation}
Summing this over all faces, we have that the total area is 
\begin{equation}
    A_{tot}/\ell^2 = F \pi - \sum_i \theta_i \; .
\end{equation}
where $F$ is the total number of faces.

From (\ref{DeltaM}), the sum of the angles around any vertex is related to the mass of the particle at that vertex by
\begin{equation}
2 \pi - \sum_{vertex} \theta_i = 8 \pi G M 
\end{equation}
Summing this over all vertices, we have 
\begin{equation}
2 \pi V - \sum_i \theta_i = 8 \pi G M_{tot} 
\end{equation}
The Euler formula relates the number of vertices, faces, and edges to the Euler characteristic $\chi = 2 - 2g$ by
\begin{equation}
    \chi = V - E + F \; .
\end{equation}
For a triangulation, we have
\begin{equation}
    E = {3 \over 2} F \; ,
\end{equation}
since each face has three edges and each edge is part of two faces. Combining all these results, we can eliminate $V$,$F$,$E$, and $\sum_i \theta_i$ to obtain
\begin{equation}
\label{Mass}
    \boxed{M_{tot} = {1 \over 4 G} ({A_{tot} \over 2 \pi \ell^2} + \chi)}
\end{equation}

Let's compare this with the results for a homogeneous and isotropic matter cosmology. For general spatial curvature $K$ and a negative cosmological constant, the Friedmann equation gives
\begin{equation}
    \left({\dot{a} \over a}\right)^2 + {K \over a^2} = 8 \pi G \rho_M - {1 \over \ell^2} \; .
\end{equation}
On the $t=0$ slice, we have $\dot{a} = 0$ and $a=1$ so this reduces to
\begin{equation}
\label{Friedslice}
     K = 8 \pi G \rho_M - {1 \over \ell^2} \; .
\end{equation}
According to the Gauss-Bonnet theorem, 
\begin{equation}
    2 \pi \chi = \int K dA = K A_{tot} \; ,
\end{equation}
so integrating (\ref{Friedslice}) over the whole spatial slice we again get precisely (\ref{Mass}). 

The evolution of the scale factor is precisely the same in both cases, so we can conclude that for this class of models, the inhomogeneities present in the exact solutions do not affect the evolution at all (there are no ``backreaction'' effects).

\subsection{Example: flat wormhole / cosmology}
\label{lattice}

\begin{figure}
 \centering
    \includegraphics[scale = 0.35]{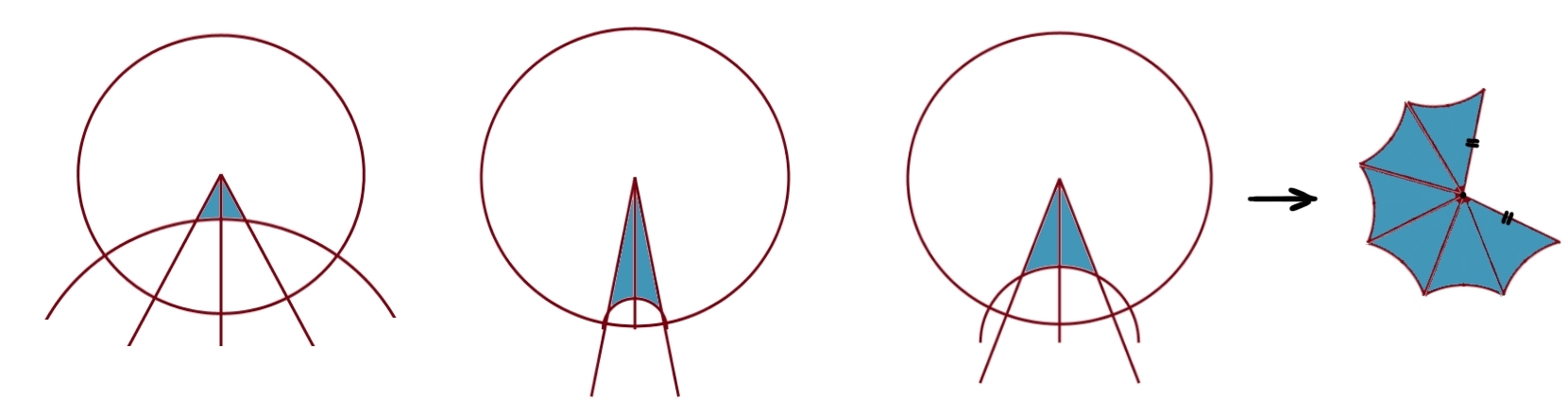}
    \caption{Construction of hyperbolic equilateral triangles with various side lengths. Right: gluing six equilateral hyperbolic triangles at a vertex. Repeating this pattern results in a triangular lattice whose geometry is locally hyperbolic away from the defects but flat on large scales.}
    \label{fig:triangular}
\end{figure} 

As a specific example, we can consider a ``flat'' cosmology, that is, a cosmology where the area of a ball (averaged over the location of the center point) increases as the square of the radius for large radius despite the fact that the local geometry away from the defects has negative curvature.

For simplicity, we can take the matter particles to be arranged in a triangular lattice, so that the $t=0$ spatial slice is obtained by gluing equilateral hyperbolic triangles. Using the hyperbolic cosine law, the equal angles are related to the side length $L$ by
\begin{equation}
    \cos(\theta) = {1 \over 2} (\tanh^2 {L \over 2} + 1)
\end{equation}
Since we always have $\theta < \pi/3$, we can glue six together at a vertex, such that we have a smooth hyperbolic geometry away from the vertex but a conical deficit angle $\delta = 2 \pi - 6 \theta$ at the vertex. We can continue the tiling, performing the same gluings at the outside vertices to give a geometry that is locally hyperbolic with a triangular lattice of conical defects. 

In this construction, the side length of each triangle is a free parameter. The average density is the same in all cases, however.  As the side length is increased, the angles of the triangles become smaller, the deficit angle at each vertex becomes larger, so more mass is required.  It is interesting to note that in the limit where the length goes to infinity, the triangle areas approach a finite value $\pi \ell^2$ while the particle masses approach $1/(4G)$, the black hole threshold. On the other hand, as the side length $L$ becomes small the conical defect angle becomes small as well ($\delta\approx \sqrt{3} L^2\ll1$) and the particles become lighter.

\subsection{Above the black hole threshhold}
\label{sec:heavy}

We can generalize our construction to the case where some of the particle masses go above the black hole threshold. Here, the conical defects are replaced by  geodesic circles with length $L$ which can be interpreted as black hole horizons. Past these circles, we can glue in a hyperbolic funnel with an asymptotic region, or more generally, we could glue the circle to another such circle in a different part of the space.
In the Lorentzian spacetime geometry, the geodesic circle becomes a black hole horizon. When we have a funnel glued in past the circle, this gives rise to a second asymptotic region of a two-sided black hole.

\subsubsection*{Example: flat cosmology with black holes}

\begin{figure}
 \centering
    \includegraphics[scale = 0.3]{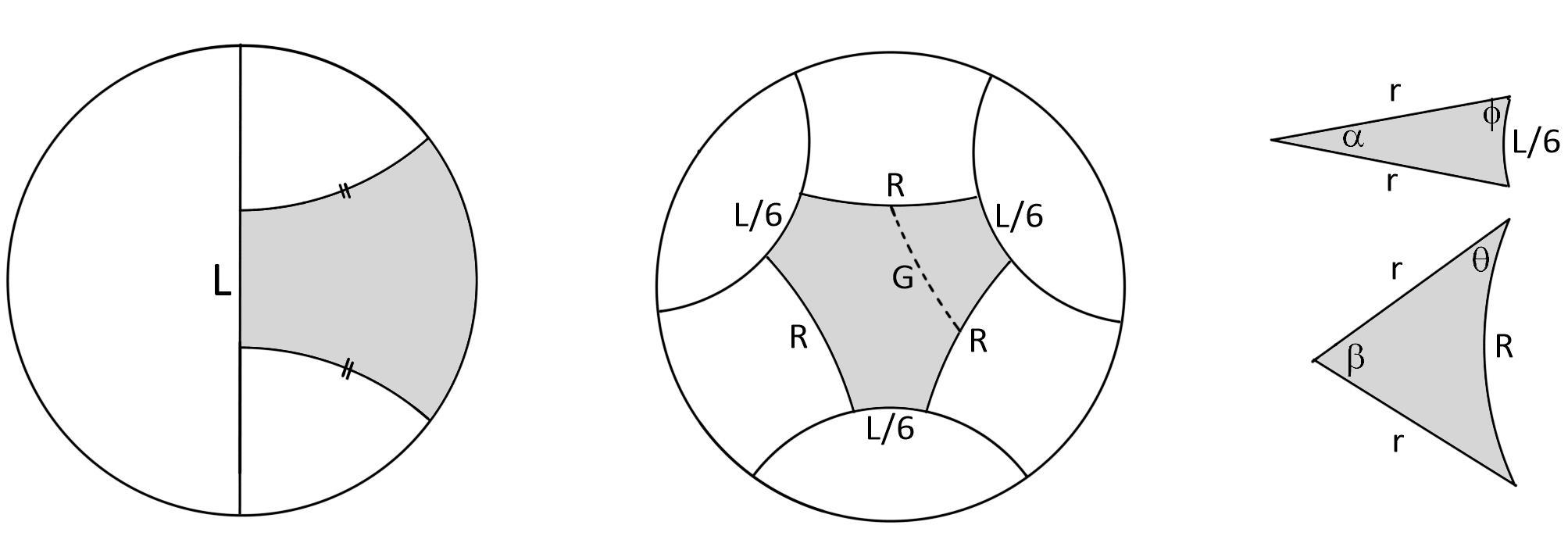}
    \caption{Left: A hyperbolic funnel ($t=0$ spatial geometry of hyperbolic black hole with horizon length $L$). Middle: hyperbolic hexagon with all angles equalling $\pi/2$. Gluing these into a hexagonal lattice along the sides of length $R$ leaves a triangular lattice of geodesic circles that can be glued on to hyperbolic funnels. This gives the spatial geometry of a flat cosmology filled with black holes. Right: the hyperbolic hexagon can be glued together from three of each of these triangles.}
    \label{fig:BHhex}
\end{figure} 

As an example, we can generalize the flat cosmology of Section \ref{lattice} to one where the triangular lattice of defects is replaced by a triangular lattice of black holes. The spatial regions between the black hole horizons can be decomposed into hyperbolic hexagons with 90 degree angles, as shown in Figure \ref{fig:BHhex}. We have a single free parameter, which can be taken as $L$, the black hole horizon length in AdS units. The geodesic distance $R$ (in AdS units) between adjacent horizons on the $t=0$ slice is fixed by the choice of $L$. In terms of the parameters in Figure \ref{fig:BHhex}, the relation is given implicitly by
\begin{eqnarray*}
    \alpha + \beta &=& 2\pi/3 \cr
    \sinh(R/2) &=& \sin(\beta/2) \sinh(r) \cr
    \sinh(L/2) &=& \sin(\alpha/2) \sinh(r) \cr
    \cosh^2(r) &=& {1 \over \tan {\alpha \over 2} \tan {\beta \over 2}}
\end{eqnarray*}
where these are derived using the hyperbolic cosine law (\ref{hypcos}) and the hyperbolic sine law $\sin(A)/\sinh(a) = \sin(B)/\sinh(b) = \sin(C)/\sinh(c)$. The geodesic distance between midpoints of the sides of length $R$ is given by
\begin{equation}
\label{geodesic}
    \sinh(G/2) = {\sqrt{3} \over 2} {1 \over \sqrt{\coth^2r \sec^2 {\beta \over 2} - 1}} \; .
\end{equation}

For all values of $L$, the area of the hyperbolic hexagon is exactly $\pi \ell^2$, so the area outside the horizons is always $2 \pi \ell^2$ per black hole.

\subsubsection*{Spacetime geometry}

\begin{figure}
 \centering
    \includegraphics[scale = 0.3]{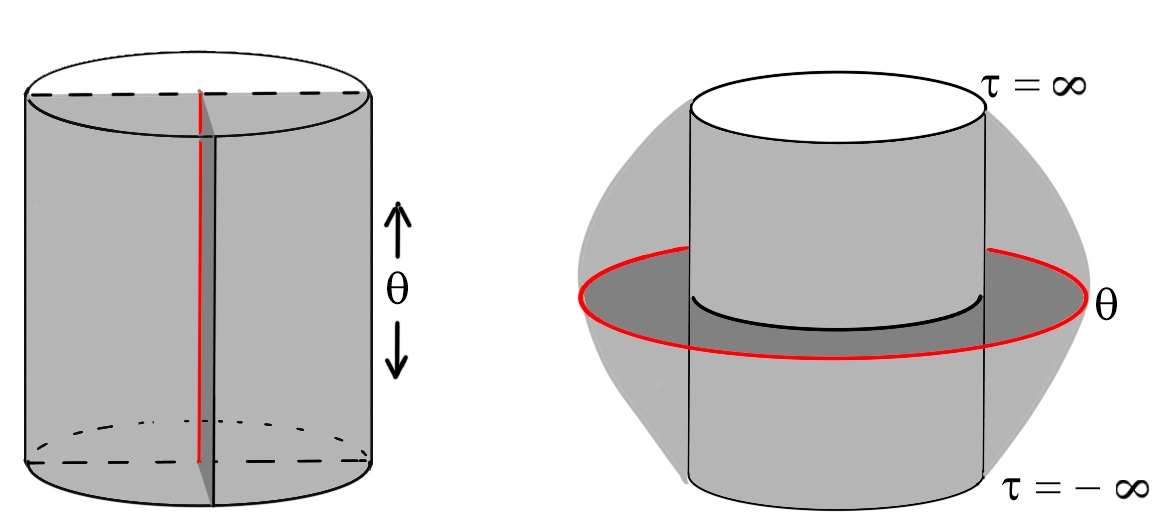}
    \caption{Left: the Euclidean geometry associated with each funnel on the $\tau = 0$ spatial slice is half the Euclidean BTZ black hole, equivalent to a portion of the global AdS cylinder, periodically identified as $\theta = \theta + 2 \pi$. The red circle represents the geodesic circle bounding the funnel. Right: Schematic representation of the same geometry, with the tubular boundary geometry shown in the middle. In both pictures, the dark shaded region represents the funnel at $\tau=0$.}
    \label{fig:Hole}
\end{figure}

As above, the spacetime geometry in the cosmology picture can be obtained from the spatial geometry $d \Sigma^2$ as
\begin{equation}
    ds^2 = -dt^2 + \cos^2(t/\ell) d \Sigma^2 \; .
\end{equation}
For each funnel region, the geometry can be extended to half of the maximally extended BTZ black hole geometry, the region $y > 0$ in the metric
\begin{equation}
    ds^2 = {1 \over \cos^2(y/\ell)}(-ds^2 + dy^2 + 
 {L^2 \over 4 \pi^2 }\cos^2(s/\ell) d \theta^2) \; .
\end{equation}
In the Euclidean geometry
\begin{equation}
\label{Euc}
    ds^2 = d \tau^2 + \cos^2(\tau/\ell) d \Sigma^2 \; ,
\end{equation}
the geometry that arises from regions  of $\Sigma$ excluding the funnels has an asymptotic geometry that has the topology of a pair of planes with a triangular lattice of disks removed. The geometry corresponding to each funnel region is equivalent to half of the Euclidean BTZ geometry whose boundary is a torus, as shown in Figure \ref{fig:Hole}. This part of the geometry has an asymptotic boundary that is a cylinder connecting $\tau = \infty$ and $\tau = - \infty$. Thus, in the full Euclidean geometry, we have an asymptotic boundary that is a pair of planes joined together by a lattice of tubes. 

A more detailed investigation of the Euclidean solutions involving black holes and the corresponding black hole cosmologies appears in \cite{Sahu:2024ccg}.

\subsection{Away from the heavy particle limit. }
\label{sec:QFT}

The conical defect solutions that we have considered provide an accurate description in a limit of large $c$ with particle masses in AdS units scaling as $\epsilon_i c$. If we consider large but finite $c$, these particles should be described explicitly in terms of quantum fields backreacting on the geometry. In this section, we consider this field theory description to understand better how this goes over to the conical defect picture in the limit of large $c$. An important point is that the defect solutions do not arise as the limit of classical solutions with scalar fields, but rather the limit of single-particle configurations for which the expectation value of the field operator vanishes.

For simplicity, we consider the case of a single defect/particle in a global AdS background. The local physics near this particle should provide a good description of the local physics near a similar particle in a cosmological/wormhole solution.

\subsubsection*{Single Particle in AdS}

A heavy massive particle in AdS${}_3$ with mass below the black hole threshhold is described by the conical defect solution 
\begin{equation}
ds^2 = \frac{1}{\cos^2(\rho)} \left( -dt^2 + d\rho^2 + \sin^2(\rho) \, d\phi^2 \right)
\end{equation}
where $\phi \in [0,2\pi - \delta)$ and $\delta$ is related to the particle mass via (\ref{DeltaM}). We are setting $\ell_{AdS}=1$ in this section. This is a rotationally-symmetric static solution, suggesting an rotationally-invariant energy eigenstate of the corresponding CFT or more generally a mixed state with density matrix commuting with the Hamiltonian and rotation generators.
If we wish to consider the description of analogous states with an explicit description in terms of fields, the simplest states with these properties are rotationally-invariant single-particle energy eigenstates. In this section, we will study the stress-energy tensor for such states in the limit where the mass becomes large.
\\
We consider a single scalar particle in global AdS${}_3$, with metric
\begin{equation}
    ds^2 = \frac{1}{\cos^2(\rho)} \left( -dt^2 + d\rho^2 + \sin^2(\rho) \, d\phi^2 \right)
\end{equation}
where $t\in\mathbb{R},\, \rho\in[0,\pi/2), \text{ and } \phi\in[0,2\pi)$. A massive scalar field in AdS satisfies the usual Klein-Gordon equation,
\begin{equation}
    \nabla^\mu \nabla_\mu \varphi + m^2 \varphi^2 = 0 \; .
\end{equation}
A general solution can be expanded in terms of a basis of solutions $\psi_{nl}$ as
\begin{equation}
\label{expandfield}
    \varphi(t,\rho,\phi) = \sum_{n,l} \left( a_{nl}\, \psi_{nl} + a_{nl}^{\dagger}\, \bar{\psi}_{nl}  \right)
\end{equation}
where each of the mode functions $\psi_{nl}$ are
\begin{equation}
    \psi_{nl}(t,\rho,\phi) = \frac{1}{N_{n,l}} e^{-iE_{nl}t + il\phi} \cos^{\Delta}(\rho) \sin^l (\rho) \, _2F_1\left(-n, \Delta + n + l, l + 1; \sin^2(\rho) \right)
\end{equation}
with energies $E_{n,l}$ given by
\begin{equation}
    E_{nl} = \Delta + 2n + l \qquad \qquad \Delta (\Delta - 2) = m^2 
\end{equation}
where the normalisation factors
\begin{equation}
    N_{n,l} = (-1)^n \sqrt{\frac{n! \Gamma(\Delta + n) \Gamma(l + 1)^2}{\Gamma(n + l + 1) \Gamma(\Delta + n + l)}}
\end{equation}
are fixed by demanding
\begin{equation}
    \langle \psi_{n,l} , \psi_{n'l'} \rangle = \int d^{2}x \, \sqrt{g}\, g^{00}\, \left[ \psi_{nl} \,(\partial_t \psi_{n'l'}^{\dagger}) - (\partial_t \psi_{nl}) \psi_{n'l'}^{\dagger} \right] = \delta_{n,n'} \delta_{l,l'} \; .
\end{equation}
In the quantum theory, (\ref{expandfield}) describes the expansion of the field operator in terms of creation and annihilation operators. The canonical quantisation conditions on $\varphi$ and $\pi = \partial L/\partial \dot\varphi = (\sin \rho/ \cos \rho)\dot{\varphi}$ give the standard commutation relations for the creation and annihilation operators:
\begin{equation}
    [\varphi(x),\pi(y)] = i (2\pi)^2 \delta^2(x-y) \implies [a_{nl}, a_{n'l'}^{\dagger}] = \delta_{n,n'} \delta_{l,l'} \; .
\end{equation}
We can now define the ground state as the state annihilated by the $a_{nl}$.
\begin{equation}
    a_{nl} \ket{0} = 0 \quad \forall n,l
\end{equation}
The other states in the Hilbert space are obtained by acting with the creation operators $a_{n,l}^\dagger$ on $\ket{0}$.
\begin{equation}
    (a_{n_1 l_1}^{\dagger}a_{n_2 l_2}^{\dagger}\dots) \ket{0} = \ket{(n_1 l_1), (n_2 l_2), \dots}
\end{equation}
Let us now look at the stress-tensor for the quantized field.
\begin{equation}
    T_{\mu\nu} = \partial_\mu \varphi \partial_\nu \varphi - \frac{g_{\mu\nu}}{2} \left( g^{\rho\sigma} \partial_\rho \varphi \partial_\sigma \varphi + m^2 \varphi^2 \right)
\end{equation}
Specifically, the $tt$-component and the $\rho\rho$-component are given as
\begin{equation}
    T_{tt} = \frac{1}{2} \left( \dot \varphi^2 + \varphi'^2 + \frac{(\partial_\phi \varphi)^2}{\sin^2 \phi} + \frac{m^2 \varphi^2}{\cos^2(\rho)} \right)
\end{equation}
\begin{equation}
    T_{\rho\rho} = \frac{1}{2} \left( \dot \varphi^2 + \varphi'^2 - \frac{(\partial_\phi \varphi)^2}{\sin^2 \phi} - \frac{m^2 \varphi^2}{\cos^2(\rho)} \right)
\end{equation}
We now compute the expectation value of the stress tensor $\langle nl| T_{\mu\nu} |nl\rangle \equiv \langle T_{\mu\nu} \rangle _{nl}$ in the single particle state $\ket{nl} = a_{nl}^{\dagger} \ket{0}$. The following identity is useful for this computation
\begin{equation}
    \langle \partial_\mu \varphi \, \partial_\nu \varphi \rangle _{nl} = (\partial_\mu \bar\psi_{nl}) \, (\partial_\nu \psi_{nl}) + (\partial_\nu \bar\psi_{nl}) \, (\partial_\mu \psi_{nl})
\end{equation}
in writing this result, we have assumed normal ordering as usual and removed the zero-point energy term. Substituting this in the previous equations, we obtain
\begin{equation}
    \langle T_{tt} \rangle _{nl} =  \abs{\partial_t \psi_{nl}}^2 + \abs{\partial_\rho \psi_{nl}}^2 + \frac{\abs{\partial_\phi \psi_{nl}}^2}{\sin^2\rho} +  \frac{m^2 \abs{\psi_{nl}}^2}{\cos^2\rho} 
\end{equation}
\begin{equation}
    \langle T_{\rho\rho} \rangle _{nl} =  \abs{\partial_t \psi_{nl}}^2 + \abs{\partial_\rho \psi_{nl}}^2 - \frac{\abs{\partial_\phi \psi_{nl}}^2}{\sin^2\rho} -  \frac{m^2 \abs{\psi_{nl}}^2}{\cos^2\rho}
\end{equation}
As a special case, consider the single particle ground state $n=0=l$. In this case,
\begin{equation}
    \psi_{00} = e^{-i\Delta t}(\cos\rho)^{\Delta}
\end{equation}
\begin{align}
    &\langle T_{tt} \rangle _{00} =  \Delta^2 \cos^{2\Delta}(\rho) + \cos^{2\Delta-2}(\rho) \, (\Delta^2 \sin^2(\rho) + m^2)  =  2\Delta (\Delta - 1) (\cos\rho)^{2\Delta-2} \\
    &\langle T_{\rho\rho} \rangle _{00} =  \Delta^2 \cos^{2\Delta}(\rho) + \cos^{2\Delta-2}(\rho)\, (\Delta^2 \sin^2(\rho) - m^2)=  2\Delta (\cos\rho)^{2\Delta-2}
\end{align}
We see that for large $\Delta$, the radial pressure is suppressed relative to the energy density, so the stress-energy tensor behaves like matter/dust. Further, the energy distribution becomes increasingly localized as we increase $\Delta$, with width $1/\sqrt{\Delta}$ since $|\psi_{00}|^2 \sim \exp(-\Delta\rho^2)$ for small $\rho$. For large $\Delta$, we approach a delta function matter source that would give rise to a conical defect when back-reaction is included.

For these single particle states, the expectation value of the field operator vanishes, so they are not well-approximated by classical field configurations. The latter are more closely related to coherent states of field. However, the coherent states are generally not energy eigenstates; as we review in Section \ref{sec:coherent} of the Appendix, these have a stress-energy tensor that oscillates in time. 

In Section \ref{sec:mixed} of the Appendix, we generalize our single particle calculation to the case of a general mixed state. We find similar results, with the stress-energy tensor of a thermal mixed state localizing in the limit of large particle mass.


\subsection*{Integrated stress-tensor and its variance}

To argue that it is a reasonable approximation to take the stress-energy tensor as a classical source for the metric, we study the integrated stress-tensor on a ball of fixed radius around the origin and look at the variance of this.
\\
We begin by defining the integrated stress-tensor operator as
\begin{equation}
    O_T = \frac{1}{2\pi}\int_{\Sigma} d^2x \,\sqrt{\gamma}\, n_\mu K^\nu \, T^{\mu}_{\nu}
\end{equation}
where $\gamma$ is the induced metric on the $t=0$ surface. $n = n_\mu dx^\mu$ is the unit normal vector and $K = \partial_t$ is the Killing vector with $\Sigma$ being a codimension-1 spacelike surface. For the case of the AdS$_3$, this integral can be then be written as
\begin{equation}
    O_T = \int \frac{d\phi}{2\pi} \int d\rho \, \frac{\sin\rho}{\cos\rho} \, T_{tt}
\end{equation}
We want to compare the expectation value and variance of the operator $O_T$ in $\psi_{00}$ state. To compute the expectation value of $O_T$ in this state we can use our result from the previous section. Plugging it in the above integral we get
\begin{equation}
    \avg{O_T}_{00} = \Delta\left( 1 - e^{-2(\Delta-1)R/l} \right)
\end{equation}
where we have perform a change of coordinates $\cos\rho = e^{-r/l}$ and taken the surface $\Sigma$ to be a disk of radius $r=R$. Note, that if we take $R\rightarrow\infty$, we just get $\Delta$ since we are in an energy eigenstate with $\Delta$ as the eigenvalue. Now, we can compute the two-point function of $O_T$ to investigate the variance.
\begin{equation}
    \begin{split}
        \avg{O_T \, O_T}_{00} &= \int_{R} \frac{d\phi_1 dr_1}{2\pi l} \int_{R} \frac{d\phi_2 dr_2}{2\pi l} \avg{T_{tt}(\phi_1, r_1) T_{tt}(\phi_2, r_2)}_{00} \\
        &= \sum_{n,l} \int_{R} \frac{d\phi_1 dr_1}{2\pi l} \int_{R}\frac{d\phi_2 dr_2}{2\pi l} \bra{00} T_{tt}(\phi_1, r_1) \ket{nl} \bra{nl} T_{tt}(\phi_2, r_2) \ket{00} \\
    \end{split}
\end{equation}
Performing the $\phi$ integrals projects onto states with $l=0$.
\begin{equation}
    \begin{split}
        \avg{O_T \, O_T}_{00} &= \sum_{n} \int_{0}^{R} \frac{dr_1}{l} \int_0^{R}\frac{dr_2}{l} \bra{00} T_{tt}(r_1) \ket{n0} \bra{n0} T_{tt}(r_2) \ket{00} \\
        &= \sum_{n} \int_{0}^{R} \frac{dr_1}{l} \int_0^{R}\frac{dr_2}{l} \bra{00} T_{tt}(r_1) \ket{n0} \bra{n0} T_{tt}(r_2) \ket{00} \\
    \end{split}
\end{equation}
The expressions $\bra{00} T_{tt}(r_1) \ket{n0}$ can be evaluated using the explicit form of wavefunctions $\psi_{nl}$ and the commutation relations.
\begin{equation}
    \begin{split}
        \bra{00} T_{tt}(r) \ket{n0} =\, & 2\Delta e^{-2(\Delta-1) r/l} (-1)^{n} [ n  (\Delta +n) e^{-2r}\left(1-e^{-2r}\right)\, _2F_1\left(1-n,n+\Delta+1,2;1-e^{-2 r}\right) \\
        &\quad\quad + \left((\Delta -1) + ne^{-2r}\right) \, _2F_1\left(-n,n+\Delta ;1;1-e^{-2 r}\right) ]
    \end{split}
\end{equation}
the integrals that appear above can then be done for each $n$ in the sum. For a fixed $R$ and taking $\Delta$ to be large the contribution from different $n$ scales as
\begin{equation}
    \int \abs{\bra{00} T_{tt} \ket{n0}}^2 \sim e^{-4(\Delta-1)R/l + \mathcal{O}(\log \Delta)} \,, \quad (n>0)
\end{equation}
which shows that the energy density in this state is localised around a region of size $l/\Delta$ and in the limit of large $\Delta$ or mass of the particle this gives narrow distribution around the center of AdS. This implies that the variance of energy distribution is small in this limit; it therefore seems reasonable to use the expectation value of the stress-tensor as a classical source. 

\section{CFT picture and interpretations}
\label{sec:CFT}

In the previous section, we have constructed exact Euclidean wormhole solutions of three-dimensional gravity with a negative cosmological constant, supported by heavy matter particles. These solutions have conformal boundaries at $\tau = \pm \infty$ that are asymptotically AdS away from the defects. In this section, we discuss the possible holographic interpretation of these solutions, that is, a microscopic construction in terms of a dual holographic CFT (or possibly a family of such CFTs).

\subsection{Heavy particles from operator insertions}

The presence of two asymptotically locally AdS boundaries suggests that the dual CFT picture should involve a pair of Euclidean CFTs. According to the standard AdS/CFT dictionary, each heavy particle in the wormhole geometry should correspond to the insertion of a pair of operators with dimension
\begin{equation}
    \Delta \approx M \ell
\end{equation}
at either end of the particle trajectory, where $M$ is the mass of the particle. So the wormhole solutions are apparently related to the partition function for a pair of Euclidean CFTs with many pairs of heavy operator insertions at corresponding points on the two CFTs.

For a wormhole with spatial geometry $\Sigma$, each CFT lives on a two-dimensional Euclidean manifold conformal to $\Sigma$. While $\Sigma$ itself is locally hyperbolic in our examples, the metric can be altered via a conformal rescaling $ds^2 \to \lambda^2 ds^2$. For the globally flat cosmology of Section \ref{lattice}, we can take the CFTs to live on Euclidean space, with a triangular lattice of operator insertions.\footnote{A Weyl transformation mapping a hyperbolic triangle (excluding the vertices) to a planar triangle can be constructed explicitly using the results of  \cite{harmer2003conformal} together with standard Schwarz-Christoffel transformations.} 

\subsection{Wormholes from ensembles of insertions}

So far, we have understood that the CFT construction corresponding to our Euclidean wormhole should
involves a pair of CFTs with operator insertions corresponding to the endpoints of the particle trajectories. Naively, this corresponds to a CFT path integral 
\begin{equation}
\label{factored}
    Z = \int [d \phi_1] e^{-S_1} \prod_a {\cal O}^{(1)}_{\alpha_a} (x_a) \int [d \phi_2] e^{-S_2} {\cal O}^{(2)}_{\alpha_a} (x_a)
\end{equation}
where $\alpha_a$ indexes which operator we are inserting. However, we are now confronted with the usual factorization puzzle \cite{Maldacena:2004rf}. This expression factorizes, while the wormhole geometry does not.
In particular, any connected correlation function between the two CFTs (obtained by inserting additional operators into the path integral above) will vanish, while such a correlation function calculated via the wormhole using the extrapolate dictionary will give a non-zero answer.

As we reviewed in the introduction, the results of \cite{Chandra_2022} show that the wormhole results have a precise interpretation in terms of CFT language. The gravitational action for the wormhole saddles matches a CFT calculation in which the products of OPE coefficients arising in the CFT correlator described above are replaced by their averages in a certain ensemble. This ensemble averaging represents a coarse-grained version of the microscopic CFT calculation, since the ensemble average results depend only on the dimensions of the operators appearing in the OPE coefficients and not on any specific details of the underlying CFT.

An interesting aspect of the correlators that correspond to our cosmological wormholes is that the correlators themselves are labeled by a significant amount of CFT data, and in passing from the classical bulk picture to the CFT, it is natural to consider an ensemble of possible microscopic expressions that could match with the bulk description. For example, when we have a heavy particle in the bulk solution, we don't know the precise quantum state of that particle, or even which precise CFT operator creates it. Thus, it is natural to associate the bulk solution with an ensemble of similar CFT correlators. This is familiar from the description of AdS black holes. When we have a black hole in AdS but are ignorant about the precise microstate, we associate that solution to a thermal density operator, the canonical ensemble of CFT energy eigenstates.  

For one of our wormholes, we might represent the ensemble of insertions as
\begin{figure}[H]
 \centering
    \includegraphics[scale = 0.3]{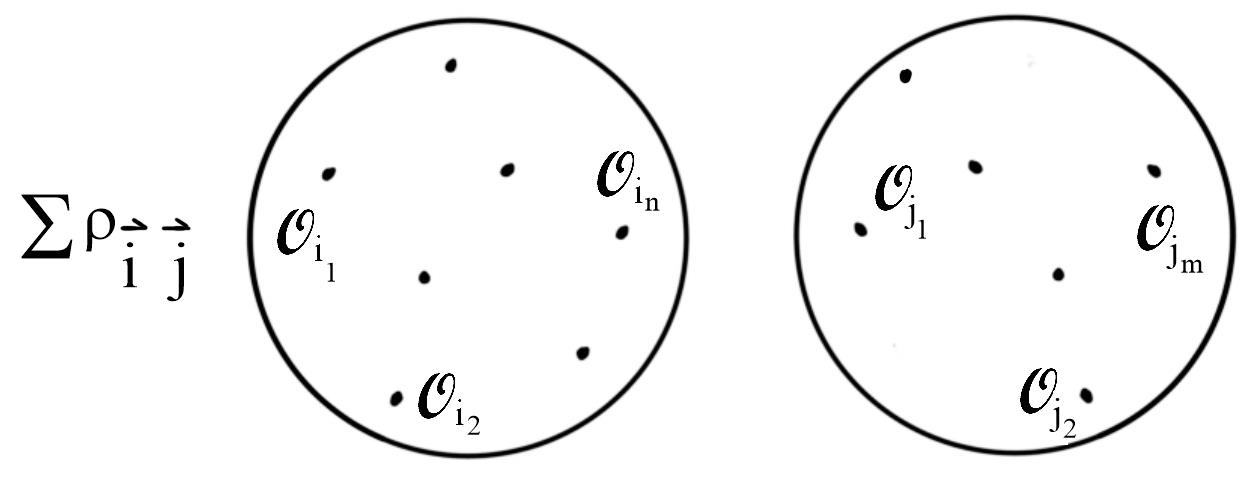} 
    \label{fig:TFD3}
\end{figure} 
\noindent
where the indices can generally indicate which operator is inserted and also the spatial location. A general mixed state bulk wavefunction for a given type of particle will involve a sum over the associated primary operator and its various descendants. It would be interesting to understand in general when such ensembles of observables give rise to a wormhole solution in the case of a single microscopic CFT without further averaging over CFT data.

\section{Sinusoidal scalar wormholes and cosmologies}
\label{sec:radiation}

In this final section, we consider a different wormhole construction where the matter comes from a collection of classical scalar fields modelling radiation. Solutions of this type were considered previously in \cite{Marolf:2021kjc}.

Again, we have explicit inhomogeneities as the scalar fields vary spatially. These inhomogeneities are in a sense what allow the solutions to exist, since at the time reflection invariant slice, the cosmological term in the Friedman equation is balanced by the spatial derivatives of the fields.

We consider a set of free scalar fields coupled to gravity in $D$ spactime dimensions. The Euclidean action is
\begin{equation}
    S = -\frac{1}{16\pi G}\int d^D x \, \sqrt{g}\, (R - 2\Lambda) \, +\,  \sum_{I=1}^{N}\frac{1}{2}\int d^D x (\,\sqrt{g}\,\, g^{\mu \nu} \partial_{\mu}\phi^I \partial_{\nu} \phi^I + m^2 \phi^I \phi^I) \; .
\end{equation}
The equations of motion derived from this action are the Klein-Gordon equation
\begin{equation}
    \label{eomscalar}
    - \nabla^{\mu}\nabla_{\mu}\phi^I  + m^2 \phi^I= 0
\end{equation}
for the scalar fields and the Einstein equation
\begin{align}
        &\hspace{2cm} G_{\mu\nu}  +\Lambda g_{\mu\nu} = \kappa T_{\mu\nu} \\
        &T_{\mu\nu} = \sum_{I} \left[ \partial_{\mu}\phi^I \partial_{\nu}\phi^I - \frac{g_{\mu\nu}}{2} (g^{\rho\sigma}\partial_{\rho}\phi^I \partial_{\sigma}\phi^I + m^2 \phi^I \phi^I) \right]
\end{align}
where we have defined $\kappa = 8\pi G$. 

As in earlier sections, we consider the FRW type ansatz 
\begin{equation}
    ds^2 = d\tau^2 + a(\tau)^2 \, d\Sigma^2
\end{equation}
for the metric of these wormholes.
We now describe different solutions for the case where the surface $\Sigma$ is either a plane $\mathbb{R}^{d}$ or a sphere $S^{d}$ and $d\Sigma^2$ is the line element on $\Sigma$. Here, we have taken $D=d+1$.

For the rest of this section, we consider planar spatial slices, so that the metric takes the following form
\begin{equation}
    ds^2 = d\tau^2 + a(\tau)^2 \sum_{i} (dx^{i})^2
\end{equation}
where $x^{i}$ with $i=1,\dots,d$ are the boundary coordinates. We generalize to the spherical case in Appendix \ref{sec:sphere}.

In these coordinates, the Einstein tensor with the cosmological constant for this metric takes a simple diagonal form
\begin{equation}
    \label{einstein}
    G_{\tau\tau} + \Lambda g_{\tau\tau} = \frac{d(d-1)}{2} \left( \frac{\dot a^2}{a^2} - \frac{1}{l^2} \right) \,, \quad G_{ii} + \Lambda g_{ii} = (d-1)a\ddot a + \frac{d-1}{2} \left( (d-2)\dot a^2 - \frac{d}{l^2} \, a^2 \right)
\end{equation}
where we have set $\Lambda = -d(d-1)/2l^2$ with $l$ as the AdS length. From this, we note that to find solutions to Einstein equation for the FRW ansatz, the stress-tensor satisfies the following properties: (1) $T_{\mu\nu}$ is diagonal, (2) $T_{\mu\nu}$ is independent of $x^i$ i.e. it has to be homogeneous, and (3) $T_{ii}$ are equal for all $i$, due to the symmetry between spatial directions. These conditions will restrict severely the kind of solutions that we can consider for our scalar fields and hence the wormholes.

The wave equation for each scalar field (\ref{eomscalar}) in this background is given by
\begin{equation}
    \ddot\phi + d\left(\frac{\dot a}{a}\right)\dot\phi + \frac{1}{a^2} \nabla^2 \phi  - m^2 \phi= 0
\end{equation} 
For an arbitrary scalar field configuration, the stress tensor for each individual scalar field with this metric ansatz is
\begin{equation}
    \begin{split}
        T_{\tau\tau} &= \frac{\dot\phi^2}{2} -  \frac{\sum_{i} (\partial_{i}\phi)^2 }{2a(\tau)^{2}} - {1 \over 2} m^2 \phi^2 \\
        T_{ii} &= -\frac{a(\tau)^2 \dot\phi^2}{2} + \frac{ (\partial_{i}\phi)^2 }{2} -  \frac{\sum_{j\ne i} (\partial_{j}\phi)^2 }{2} - {a^2(\tau) \over 2} m^2 \phi^2\\
        T_{\mu\nu} &= \partial_{\mu}\phi \, \partial_{\nu}\phi \,, \quad (\mu\ne\nu)
    \end{split}
\end{equation}
The full stress tensor is given by adding such terms for each scalar. It is clear from these expressions that we will need to fine-tune the scalar field profiles such that the off-diagonal terms vanish and that we get something homogeneous. 

\subsubsection*{Time Independent Scalars}

A simple class of solutions for the planar boundary problem is where the the scalar field configurations don't change with (Euclidean) time $\tau$. In this case, the equations $G_{\tau i} = g_{\tau i} + \kappa T_{\tau i}$ are trivially true. But we also want $T_{ij}=0$ for $i\ne j$. If we had only one scalar then this condition implies that $\phi$ is a function of only one of the $x^i$ coordinate. However, this is inconsistent since the form (\ref{einstein}) requires all $T_{ii}$ to be equal. We see that a single real scalar field can never support such a wormhole solution. However, it is easy to see that with $d$ massless scalars, one can indeed find a solution. Choosing the field profiles as
\begin{equation}
    \phi^I(\tau,x,y) = x^I \,,\quad  (I=1,\dots,d)
\end{equation}
These clearly obey the scalar equation of motion. The stress-tensor for such a field configuration is
\begin{equation}
    T_{\tau\tau} = -\frac{d}{2a^2} \,, \quad T_{ii} = -\frac{d-2}{2}
\end{equation}
all the other components of the stress tensor vanish. The Einstein equations in this case reduce to
\begin{equation}
    \quad \dot a(\tau)^2 = \frac{a(\tau)^2}{l^2} - \frac{\kappa}{d-1}  
\end{equation}
\begin{equation}
    (d-1)a\ddot a + \frac{d-1}{2} \left( (d-2)\dot a^2 - \frac{d}{l^2} \, a^2 \right) = -\frac{\kappa(d-2)}{2}
\end{equation}
These are not independent and taking a time derivative of the first leads to the second. The solution $a(\tau)$ satisfying the above equations is
\begin{equation}
    a(\tau) = \sqrt{\frac{\kappa l^2}{d-1}} \, \cosh(\frac{\tau - \tau_0}{l})
\end{equation}
$\tau_0$ is the turning point for $a(\tau)$ or the throat of the wormhole. 

\subsubsection*{Time Dependent Scalars}
We can also consider the case where the fields are allowed to have a non-trivial time dependence. We can consider the following form for the scalar fields
\begin{equation}
    \phi^I(\tau,x^i) = h(\tau)f^{I}(x^i)
\end{equation}
where for now $h(\tau)$ and $f^{I}(x^i)$ are arbitrary function that depend on time and spatial coordinates only. Note that the time dependent part is chosen to be the same for all scalar fields whereas the spatial part can be different. We can fix the function $f^I$ by looking at the constraints described above.
\begin{equation}
    T_{\tau i} = \frac{h\dot{h}}{2}\sum_I \partial_i (f^{I})^2 = 0 
\end{equation}
where we set it to $0$ to satisfy Einstein equations. For $h(\tau)$ to be non-trivial function of time, this imposes the constrain that $\sum_I (f^{I})^2$ is a constant i.e. the $f^I$ are a parametrization of the sphere. The condition that all $T_{ii}$ components are equal gives
\begin{equation}
    \sum_I (\partial_i f^{I})^2 = \sum_I (\partial_j f^{I})^2
\end{equation}
finally, imposing $T_{ij}=0\,, (i\ne j)$ 
\be
    \sum_I \partial_i f^{I}\, \partial_j f^{I} = 0
\ee
If we consider field configurations such that each $f^I$ only depend on one of the coordinate $x^i$, then the last condition is easily satisfied. The other restrictions can be satisfied by choosing $2d$ fields such that $f^I = (\cos(kx^1), \sin(kx^1), \cos(kx^1), \sin(kx^1), \dots)$. The scalar fields are then
\begin{equation}
    \begin{split}
    \phi^I_1(\tau,x^i) = h(\tau)\cos(k x^I) \,\,,\quad \phi^I_2(\tau,x^i) = h(\tau)\sin(k x^I) \,,\quad (I=1,\dots,d)
    \end{split}
\end{equation}
Equivalently, we can have $d$ complex scalar fields $\Psi^I$ with
\begin{equation}
    \Psi^I = \frac{h(t)}{\sqrt{2}}\, e^{ikx^I}  \; .
\end{equation}
The time-dependence $h(\tau)$ along with the scale factor are determined by solving equation of motion. The stress tensor for these scalar field profiles is
\begin{equation}
    T_{\tau\tau} = \frac{d}{2}\left( \dot h^2-\frac{k^2 h^2}{a^2} \right) \,, \quad T_{ii} = -\frac{d}{2} a^2 \dot h^2 - \frac{d-2}{2}k^2 h^2
\end{equation}
all other components vanish.
Using these, the equation of motions for the scalar and the scale factor are
\begin{equation}
    \ddot h + d\left(\frac{\dot a}{a}\right)\dot h - \left( m^2 + \frac{k^2}{a^2} \right) h = 0
\end{equation}
\begin{equation}
    \left(\frac{\dot a}{a} \right)^2  = \frac{\kappa}{(d-1)}\left( \dot h^2 - \left( m^2 + \frac{k^2}{a^2} \right) h^2 \right) + \frac{1}{l^2}
\end{equation}
where we have introduced  the AdS radius $\ell$ back in the equations. The equation from other components of Einstein equations are not independent but can be derived from these. 

These equations can be simplified by rescaling
\begin{equation}
    a =  l a_s \,\,, \quad  h = \sqrt{\frac{d-1}{\kappa}} \, h_s \,, \quad \tau = l \tau_s
\end{equation}
which gives
\begin{equation}
    \ddot h_s + d\left(\frac{\dot a_s}{a_s}\right)\dot h_s - \left( m^2 l^2 + \frac{k^2}{a_s^2} \right) h_s = 0 
\end{equation}
\begin{equation}
    \left(\frac{\dot a_s}{a_s} \right)^2 =  \dot h_s^2-\left( m^2 l^2 + \frac{k^2}{a_s^2} \right) h_s^2 + 1
\end{equation}
the dot now represent a time derivative with the rescaled coordinate $\tau_s$. Now we can look for solutions such that $a_s(\tau_s)$ and $h_s(\tau_s)$ are even functions in $\tau_s$ around the throat of the wormhole at $\tau_s=0$ 
with the initial conditions
\begin{equation}
    a_s(0) = 1 \,,\quad \dot a_s(0) = 0  \,, \quad h_s(0) = \pm \frac{1}{\sqrt{m^2l^2 + k^2}} \,, \quad \dot h_s(0) = 0
\end{equation}

To find the numerical solution, we can also use the 2nd order differential equation
\begin{equation}
    \frac{\ddot a_s}{a_s} + \frac{(d-2)}{2}\left(\frac{\dot a_s}{a_s} \right)^2 = \frac{d}{2}\left(1 - \dot h^2 - \left( (ml)^2 + \frac{(d-2)k^2}{a_s^2 d} \right) h^2 \right)
\end{equation}
that follows from the others.

\begin{figure}
 \centering
    \includegraphics[scale = 0.3]{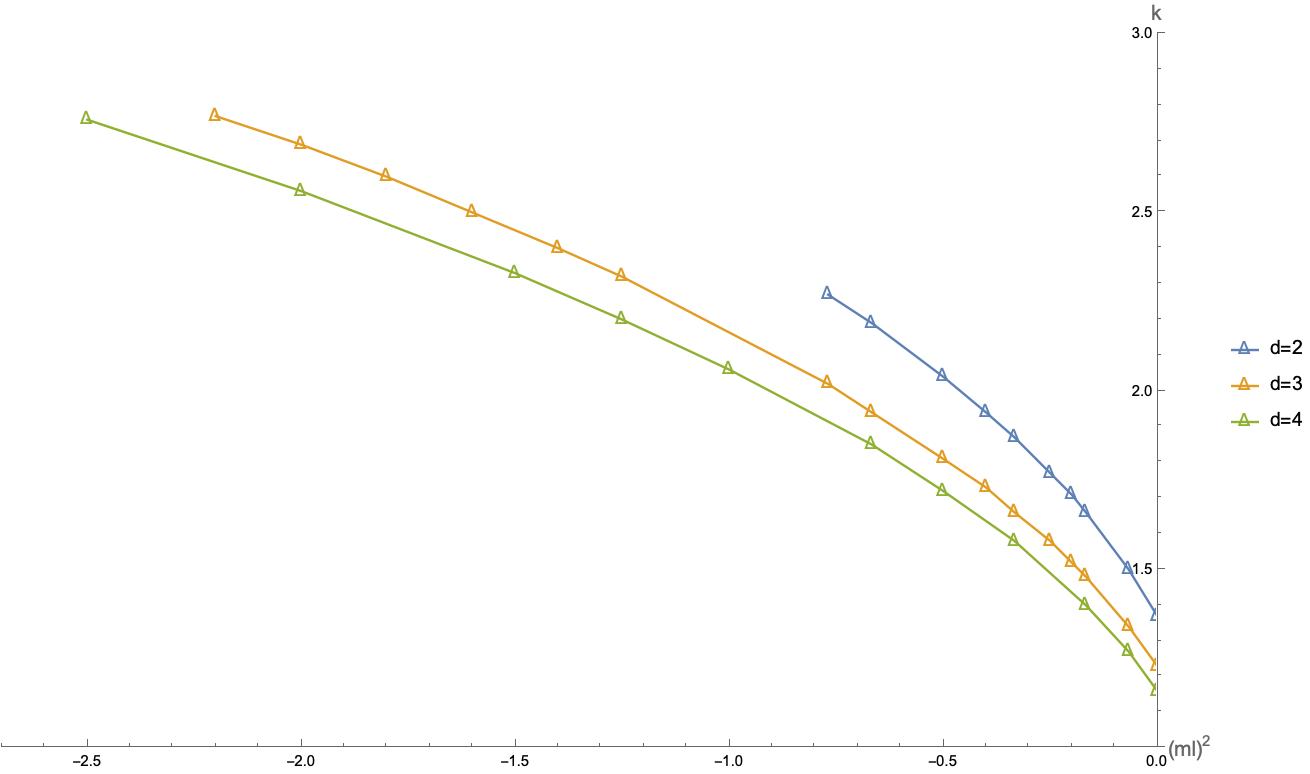}
    \caption{The figure shows the maximum allowed valued of the wavenumber $k$ as a function of the mass in AdS units $(ml)^2$ for which the wormhole solution exists.}
    \label{fig:planar_wh_k_vs_ml}
\end{figure}

There exists solution to these differential equations with the above boundary conditions for $d\ge2$. In the figure \ref{fig:planar_wh_1} below, we show a plot of these.
\begin{figure}
 \centering
    \includegraphics[scale = 0.23]{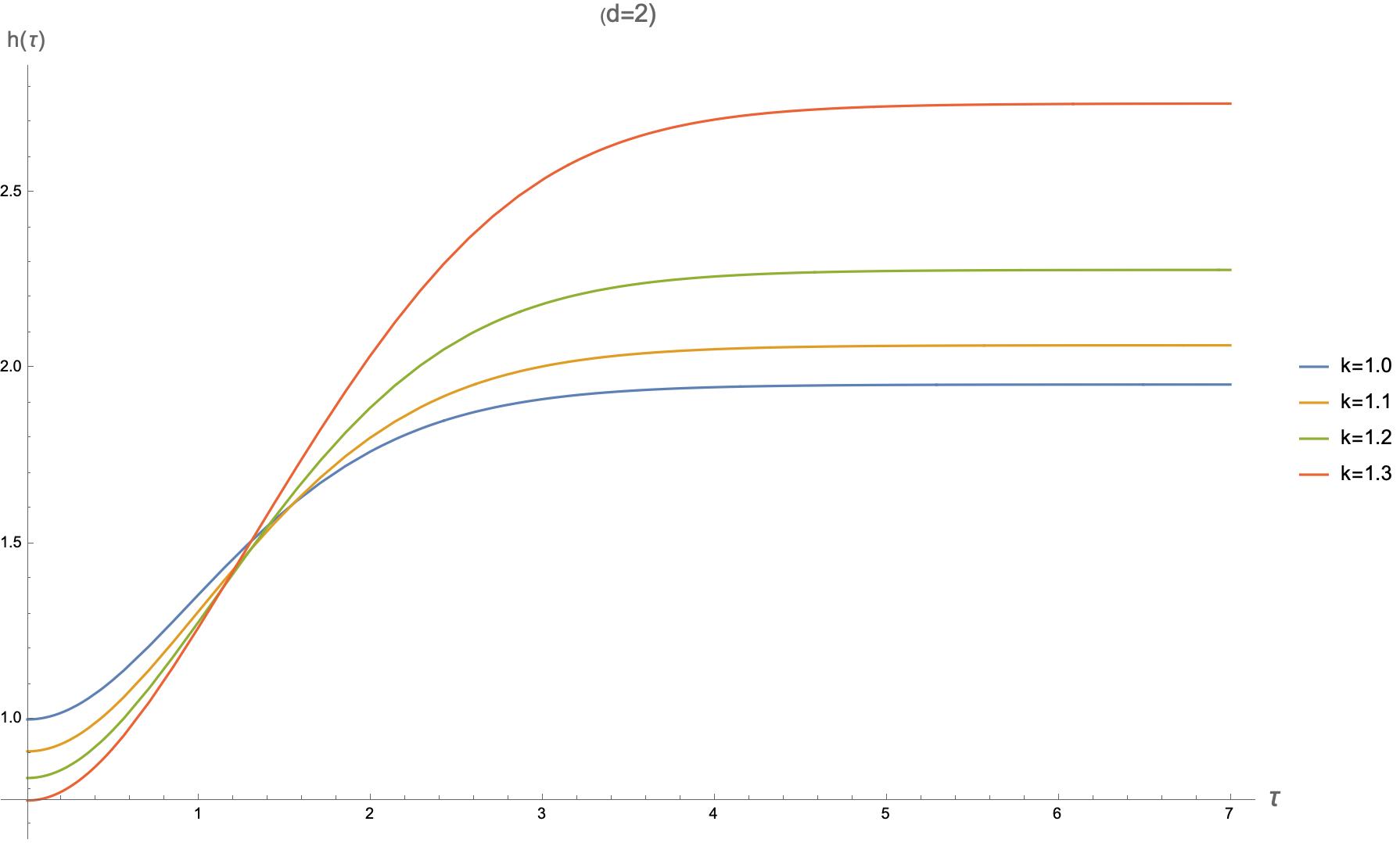} \hspace{0.5cm} \includegraphics[scale = 0.23]{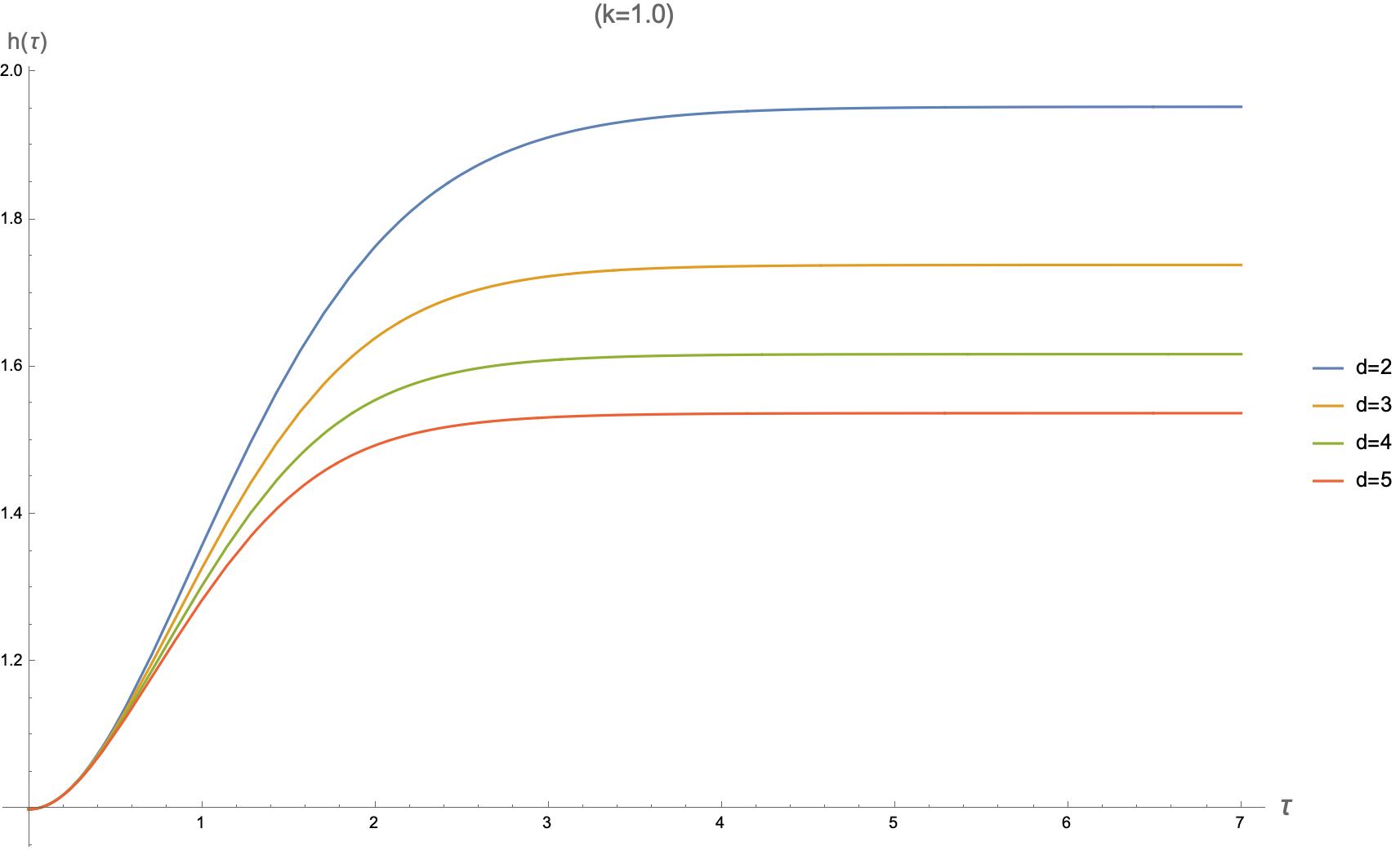}
    \caption{The time-dependence of the scalar field for planar solutions in the massless case. The plots compare the solutions for different wave number and dimensions. Since this solutions are even functions in $\tau$, we show the profile only for the region $\tau\in[0,\infty)$.}
    \label{fig:planar_wh_1}
\end{figure} 
From the numerical solutions to the differential equations, we observe that the solutions exist only below a certain maximum wave number $k_c$ or above a certain minimum wavelength of order the AdS scale. For instance, in $d=2$, we require 
\begin{equation}
    k \ell < k_c \approx 1.37
\end{equation}
while fore $d=3$, we find $k_c \approx 1.23$. In Figure (\ref{fig:planar_wh_k_vs_ml}), we show the behaviour of the critical wave number as we turn on particle mass. Note that in the plot we consider $m^2$ to be negative such that $-\frac{d^2}{4} \le (ml)^2 \le 0$ which is consistent with the Breitenlohner-Freedman bound in $(d+1)$-dimensional AdS spacetime.

For $k \ll 1$, we find that $h$ is approximately constant in time with $a(\tau) \approx \cosh(\tau_s)$. This reproduces the behavior in the constant spatial gradient solutions above.

The solutions describe here can be generalized to the case of spherical spatial slices, as we describe in Appendix \ref{sec:sphere}.

\subsection{Cosmology version}

We can consider the analogous cosmological solutions, related to these wormboles by analytic continuation of the time direction.
Here, the equations become
\begin{equation}
\label{cosmo1}
    \ddot{h} + d{\dot{a} \over a}\dot{h} + (m^2 + {k^2 \over a^2}) h = 0
\end{equation}
and
\begin{equation}
\label{cosmo2}
    \left({\dot{a} \over a}\right)^2 = -1 + \dot{h}^2 + \left(m^2 + {k^2 \over a^2}\right) h^2 \; .
\end{equation}
For the cosmological solutions, we find qualitatively similar behavior for all $k$ with a big-bang / big crunch scale factor. For the masseless case, we find that the solution behaves as 
\begin{equation}
    a(t) \to \cos(t) \qquad k \ll 1 \qquad \qquad a(t) \to \sqrt{\cos(2t)}  \qquad k \gg 1 \; .
\end{equation}
for small and large $k$.
We can understand the latter limit via a WKB solution as we describe in Appendix \ref{sec:WKB}. In this case, we just reproduce the standard solution with $\Lambda < 0$ and radiation.

A lesson here is that while the approximate scale factor obtained by modeling the scalar fields as a perfect fluid of radiation analytically continues to a wormhole solution, the exact solution only analytically continues to a wormhole for $k$ below some critical value.

\section{Discussion}

In this paper, we have considered Euclidean AdS wormhole solutions supported by ordinary matter and/or radiation, motivated by the fact that the corresponding Lorentzian cosmological solutions are ubiquitous. 

We have described a large family of such wormhole solutions in three-dimensional gravity with heavy matter particles. We found classical solutions working in the limit of large $c$ with particles whose masses are of order $c$, but we also argued that these solutions should persist for large but finite $c$. Here, it is essential to treat the matter quantum mechanically, since the heavy particles we are trying to describe are not described by classical field configurations. The field operator in the quantum description has a large variance, but the variance in the associated stress-energy tensor is small, so it makes sense to describe the geometry with a metric that has small variance. 

For Lorentzian cosmological solutions, there there is no significant qualitative difference passing from three to higher dimensions, so we expect that inhomogeneous Euclidean wormhole solutions supported by matter exist in these higher-dimensional cases as well. However, studying the solutions explicitly would likely require numerics as they will no longer be locally AdS away from the particles. 

In the context of holography, the Euclidean wormholes we consider correspond to a pair of Euclidean CFTs with a large number of pairs of operator insertions. It is natural to consider an ensemble of such insertions, corresponding to microscopically different configurations that have the same coarse-grained bulk interpretation. In the Lorentzian picture, this introduces entropy into the cosmology. In some cases with sufficient entropy, the wormhole solution can give the dominant contribution to the path integral (see \cite{Sahu:2024ccg} for an example and additional discussion). 

It will be interesting to further explore the possibilities for cosmological physics that arise from analytically continuing the types of wormhole solutions that we describe. In general, these will be mixed state cosmologies when we have an ensemble of operator insertions in the Euclidean picture. For the cosmology to arise as the dominant part of the Lorentzian bulk state, the amount of correlation in the Euclidean construction must be large. A preliminary investigation into these issues appears in \cite{Sahu:2024ccg}.

\section*{Acknowledgements}

We would like to thank Abhisek Sahu for helpful discussions.
MVR acknowledges support from the National Science and Engineering Research Council of Canada
(NSERC) and the Simons foundation via a Simons Investigator Award. 

\appendix

\section{Scalar particle stress-energy tensor calculations}

\subsection{Coherent State}
\label{sec:coherent}
We now compare how these results change for the case of a coherent state. Let us consider the coherent state for the $a_{nl}^\dagger$ mode.
\begin{equation}
    \ket{z;n,l} = e^{-\abs{z}^2/2} e^{z a_{nl}^{\dagger}}\ket{0} = e^{-\abs{z}^2/2} \sum_{m\ge0} \frac{z^m}{\sqrt{m!}} \ket{m} \,, \quad (z\in\mathbb{C})
\end{equation}
where $m$ labels the particle number for the $a_{nl}^{\dagger}a_{nl}$ operator. The expectation value of the stress tensor in this state is given by
\begin{equation}
    \begin{split}
    \langle T_{\mu\nu} \rangle _{z;n,l} &= \bra{z;n,l} T_{\mu\nu} \ket{z;n,l} \\
    \end{split}
\end{equation}
this expression contains sums of the type
\begin{align}
    \bra{m_1; n} a_{n_1} a_{n_2} \ket{m_2; n} \rightarrow \bra{m_1; n} a_{n} a_{n} \ket{m_2; n} &= \sqrt{m_2(m_2-1)} \, \delta_{m_1,m_2-2} \\
    \bra{m_1; n} a_{n_1}^{\dagger} a_{n_2}^{\dagger} \ket{m_2; n} \rightarrow \bra{m_1; n} a_{n}^{\dagger} a_{n}^{\dagger} \ket{m_2; n} &= \sqrt{(m_2+1)(m_2+2)} \, \delta_{m_1,m_2+2} \\
    \bra{m_1; n} a_{n_1}^{\dagger} a_{n_2} \ket{m_2; n} \rightarrow \bra{m_1; n} a_{n}^{\dagger} a_{n} \ket{m_2; n} &= m_2 \, \delta_{m_1,m_2}
\end{align}
We have then 
\begin{equation}
    \begin{split}
        \bra{z;n,l} \partial_i \varphi \, \partial_j \varphi \ket{z;n,l} &= z^2 (\partial_i \psi_{nl}) \, (\partial_j \psi_{nl}) + \bar{z} \, (\partial_i \bar\psi_{nl}) \, (\partial_j \bar\psi_{nl}) + z \bar{z} \, [(\partial_i \psi_{nl}) \, (\partial_j \bar\psi_{nl}) + (\partial_i \bar\psi_{nl}) \, (\partial_j \psi_{nl})] \\
        &= (z \partial_i \psi_{nl} + \bar{z} \, \partial_i \bar{\psi}_{nl})(z \partial_j \psi_{nl} + \bar{z} \, \partial_j \bar{\psi}_{nl})
    \end{split}
\end{equation}
Using this we can compute the expectation value of the stress tensor. As before, we consider the case where $n=0=l$.
\begin{align}
    \langle \dot \varphi^2 \rangle _{z;00} &= -\Delta^2 \cos^{2\Delta}(\rho) (ze^{-i\Delta t} - \bar{z} e^{i\Delta t})^2 \\
    \langle  \varphi'^2 \rangle _{z;00} &= \Delta^2 \cos^{2\Delta-2}(\rho) \sin^{2}(\rho) (ze^{-i\Delta t} + \bar{z} e^{i\Delta t})^2 \\
    m^2\langle  \varphi^2 \rangle _{z;00} &= m^2 \cos^{2\Delta}(\rho) (ze^{-i\Delta t} + \bar{z} e^{i\Delta t})^2
\end{align}
The $tt$-component and the $\rho\rho$-component are
\begin{equation}
    \begin{split}
        \langle T_{tt} \rangle _{z;00} &= \frac{1}{2} \left( -\Delta^2 \cos^{2\Delta}(\rho)(ze^{-i\Delta t} - \bar{z} e^{i\Delta t})^2 + \cos^{2\Delta-2}(\rho) \, (\Delta^2 \sin^2(\rho) + m^2)(ze^{-i\Delta t} + \bar{z} e^{i\Delta t})^2 \right) \\
        &= 2\abs{z}^2 \left( \Delta^2 \cos^{2\Delta}(\rho) \sin^2(\alpha_z - \Delta t) + \cos^{2\Delta-2}(\rho) \, (\Delta^2 \sin^2(\rho) + m^2)\cos^2(\alpha_z - \Delta t) \right) \\
    \end{split}
\end{equation}
\begin{equation}
    \begin{split}
        \langle T_{\rho\rho} \rangle _{z;00} &= 2\abs{z}^2 \left( \Delta^2 \cos^{2\Delta}(\rho) \sin^2(\alpha_z - \Delta t) + \cos^{2\Delta-2}(\rho) \, (\Delta^2 \sin^2(\rho) - m^2)\cos^2(\alpha_z - \Delta t) \right) \\
    \end{split}
\end{equation}
where we rewrite $z=\abs{z}e^{i\alpha_z}$. In the limits $\rho\rightarrow 0$ we get
\begin{equation}
    \langle T_{tt} \rangle _{z;00} = 2\abs{z}^2 (\Delta^2 \sin^2(\alpha_z - \Delta t) + m^2 \cos^2(\alpha_z - \Delta t))
\end{equation}
\begin{equation}
    \langle T_{\rho\rho} \rangle _{z;00} = 2\abs{z}^2 (\Delta^2 \sin^2(\alpha_z - \Delta t) - m^2 \cos^2(\alpha_z - \Delta t))
\end{equation}
this expectation values oscillate with a periodicity $\pi/\Delta$. For $\Delta\rightarrow\infty$, this functions oscillate rapidly. If we perform a time average, we get similar expression as before
\begin{align}
    &\langle T_{tt} \rangle  _{z;00} \rightarrow \frac{\abs{z}^2}{\Delta}\left( \Delta^2 + m^2 \right) = 2\abs{z}^2 (\Delta-1) \\
    &\langle T_{\rho\rho} \rangle_{z;00} \rightarrow \frac{\abs{z}^2}{\Delta} \left( \Delta^2 - m^2 \right) = 2\abs{z}^2
\end{align}
In this case, we see that the $T_{\rho\rho}$ component stays finite whereas the $T_{tt}$ component diverges. This is similar to the stress tensor of a perfect fluid with only energy density and no pressure terms.
\begin{equation}
    T_{\mu\nu}^{\text{fluid}} = \rho_E \, U_{\mu}U_{\nu} \,, \quad (U_\mu \propto (1,0,0) )
\end{equation}
where $\rho_E$ denotes the energy density.

\subsection{Mixed states}
\label{sec:mixed}
Instead of a pure state, we can also consider evaluating the expectation value of a stress tensor in a mixed state.
\begin{equation}
    \rho = \sum_{n,l} c_{n,l} \ket{n,l}\bra{n,l} \,, \quad \left(\sum_{n,l} c_{n,l} = 1\right)
\end{equation}
The expectation value of the stress tensor in the mixed state is then calculated using the trace.
\begin{equation}
    \langle T_{\mu\nu} \rangle_{\rho} = \text{Tr}[\,\rho\, T_{\mu\nu}] = \sum_{n,l} c_{n,l} \langle T_{\mu\nu} \rangle _{nl}
\end{equation}
where $\langle T_{\mu\nu} \rangle _{nl}$ can be computed using the $\psi_{n,l}$ and the relevant $tt$ and $\rho\rho$ components are
\begin{equation}
    \begin{split}
            \langle T_{tt} \rangle _{nl} = &\frac{(\cos \rho )^{2 \Delta -2} (\sin \rho )^{2 l-2}}{N_{n,l}^2} [ 4 n^2 \cos ^4(\rho ) \, f_{1}(\rho)^2 - 2 n \cos ^2(\rho ) (l + 2n -\Delta +\cos (2 \rho ) (\Delta + 2n + l)) \, f_1(\rho) f_2(\rho) \,  \\
            &+ \left( \Delta(\Delta -1) +l^2+2 l n+2 n^2  +\cos (2 \rho ) \left( - \Delta(\Delta -1) +l^2+2 l n+2 n^2\right)\right) \, f_2(\rho)^2 ]
    \end{split}
\end{equation}
\begin{equation}
    \begin{split}
            \langle T_{\rho\rho} \rangle _{nl} = & \frac{(\cos \rho )^{2 \Delta -2} (\sin \rho ) ^{2 l-2}}{N_{n,l}^2} [ 4 n^2 \cos ^4(\rho ) \, f_{1}(\rho)^2 - 2 n \cos ^2(\rho ) (l + 2n - \Delta +\cos (2 \rho ) (\Delta + 2n + l)) \, f_1(\rho) f_2(\rho) \,  \\
            &+ \left( 2 l n + 2 n^2 + \Delta  +\cos (2 \rho ) \left( 2 l n + 2 n^2 - \Delta \right)\right) \, f_2(\rho)^2 ]
    \end{split}
\end{equation}
\begin{align}
    f_1(\rho) &= {}_2F_1\left(1-n,l+n+\Delta ;l+1;\sin ^2(\rho )\right) \\
    f_2(\rho) &= {}_2F_1\left(-n,l+n+\Delta ;l+1;\sin ^2(\rho )\right)]
\end{align}
In the limit of $\rho\rightarrow0$, we get
\begin{equation}
    \langle T_{tt} \rangle _{nl} = 
    \begin{cases}
    \Delta(\Delta-1)+2n(\Delta+n) \,, \quad l=0 \\
        \frac{l^2}{N_{n,l}^2} \rho^{2l-2} \,, \quad l\ne0 
    \end{cases}
\end{equation}
\begin{equation}
    \langle T_{\rho\rho} \rangle _{nl} = \begin{cases}
        \Delta + 2n(n+\Delta) \,, \quad l=0 \\
        \frac{ \Delta(2n+l+1) + 2n(n+l)}{(l+1)N_{n,l}^2}\rho^{2l} \,, \quad l\ne0 
    \end{cases}
\end{equation}
The only non-zero contribution in this limit is from the $l=0$ and $l=1$ states.
\begin{equation}
    \begin{split}
        \langle T_{tt} \rangle_{\rho \rightarrow0} &= \sum_{n} c_{n,0}\left[ \Delta(\Delta-1)+2n(\Delta+n)\right] + \sum_{n} c_{n,1} \frac{1}{N_{n,1}^2}  \\
    \end{split}
\end{equation}
\begin{equation}
    \begin{split}
        \langle T_{\rho\rho} \rangle_{\rho\rightarrow0} &= \sum_{n} c_{n,0}\left[ \Delta + 2n(n+\Delta)  \right] \\ 
    \end{split}
\end{equation}
We can now consider a particular choice for the density matrix. Let us choose $c_{n,l} = \exp(-\beta E_{n,l})/Z(\beta)$ i.e. the thermal density matrix.
\begin{equation}
    Z(\beta) = \sum_{n,l} \exp(-\beta (\Delta + 2n + l)) = \frac{e^{-\beta\Delta}}{(1-e^{-\beta})^2}
\end{equation}
\begin{equation}
    \sum_{n} c_{n,0} = (1-e^{-\beta}) \,,\quad \sum_{n} n\, c_{n,0} = \frac{e^{-\beta}(1-e^{-\beta})}{1-e^{-2\beta}} \,, \quad \sum_{n} n^2\, c_{n,0} = \frac{e^{-4\beta}(1+e^{-2\beta})}{(1-e^{-2\beta})(1+e^{-\beta})}
\end{equation}
Using these,
\begin{equation}
     \langle T_{tt} \rangle_{\rho \rightarrow0} =  \Delta(\Delta-1)(1-e^{-\beta}) +  \frac{2\Delta e^{-\beta}(1-e^{-\beta})}{1-e^{-2\beta}} + \frac{2e^{-4\beta}(1+e^{-2\beta})}{(1-e^{-2\beta})(1+e^{-\beta})} + \frac{e^{-\beta}(\Delta - (\Delta-2)e^{-2\beta})}{(1-e^{-2\beta})(1+e^{-\beta})}
\end{equation}
\begin{equation}
     \langle T_{\rho\rho} \rangle_{\rho \rightarrow0} = \Delta(1-e^{-\beta}) +  \frac{2\Delta e^{-\beta}(1-e^{-\beta})}{1-e^{-2\beta}} + \frac{2e^{-4\beta}(1+e^{-2\beta})}{(1-e^{-2\beta})(1+e^{-\beta})}
\end{equation}

\begin{figure}
 \centering
    \includegraphics[scale = 0.2]{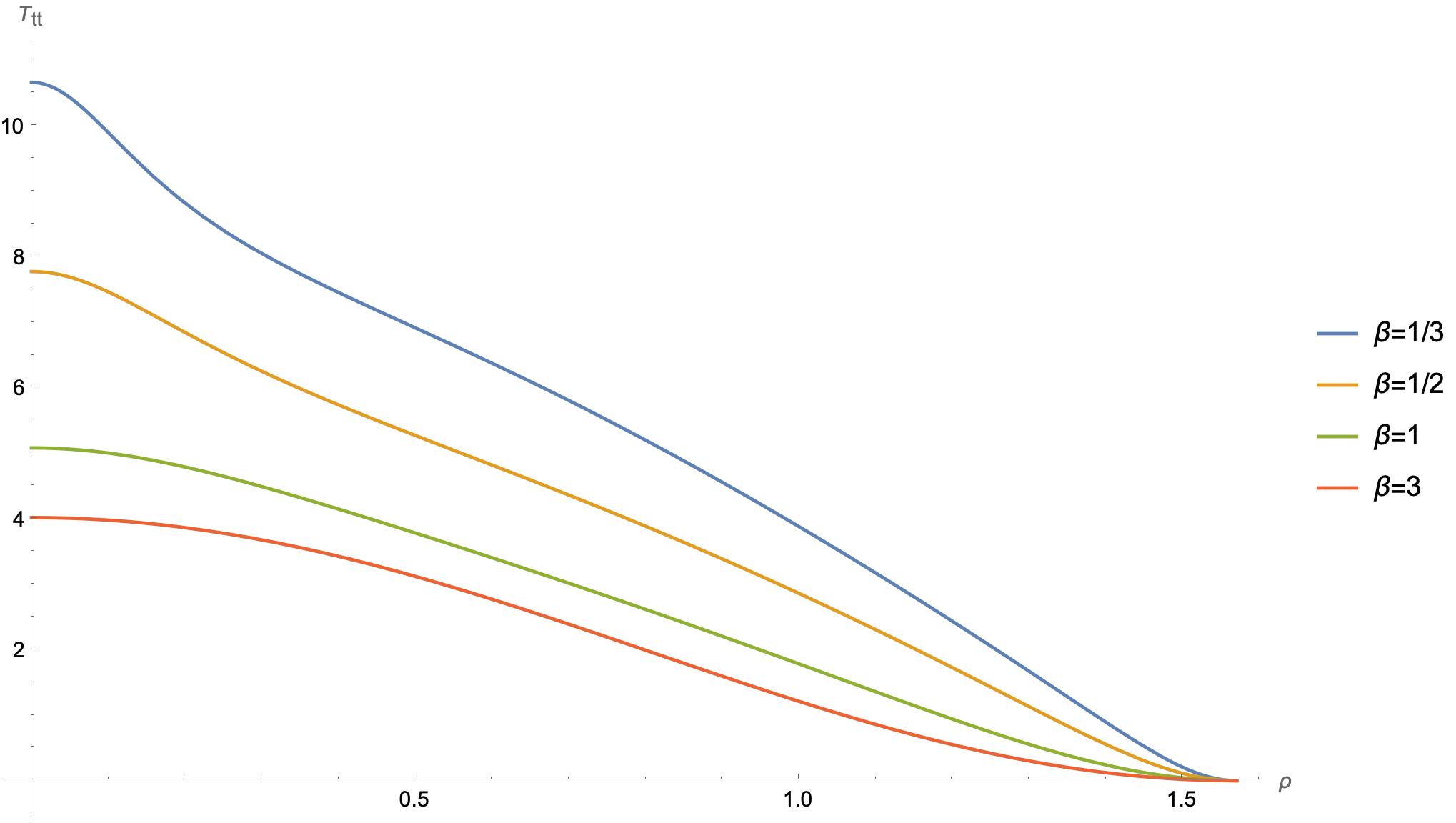} \hspace{0.5cm} \includegraphics[scale = 0.2]{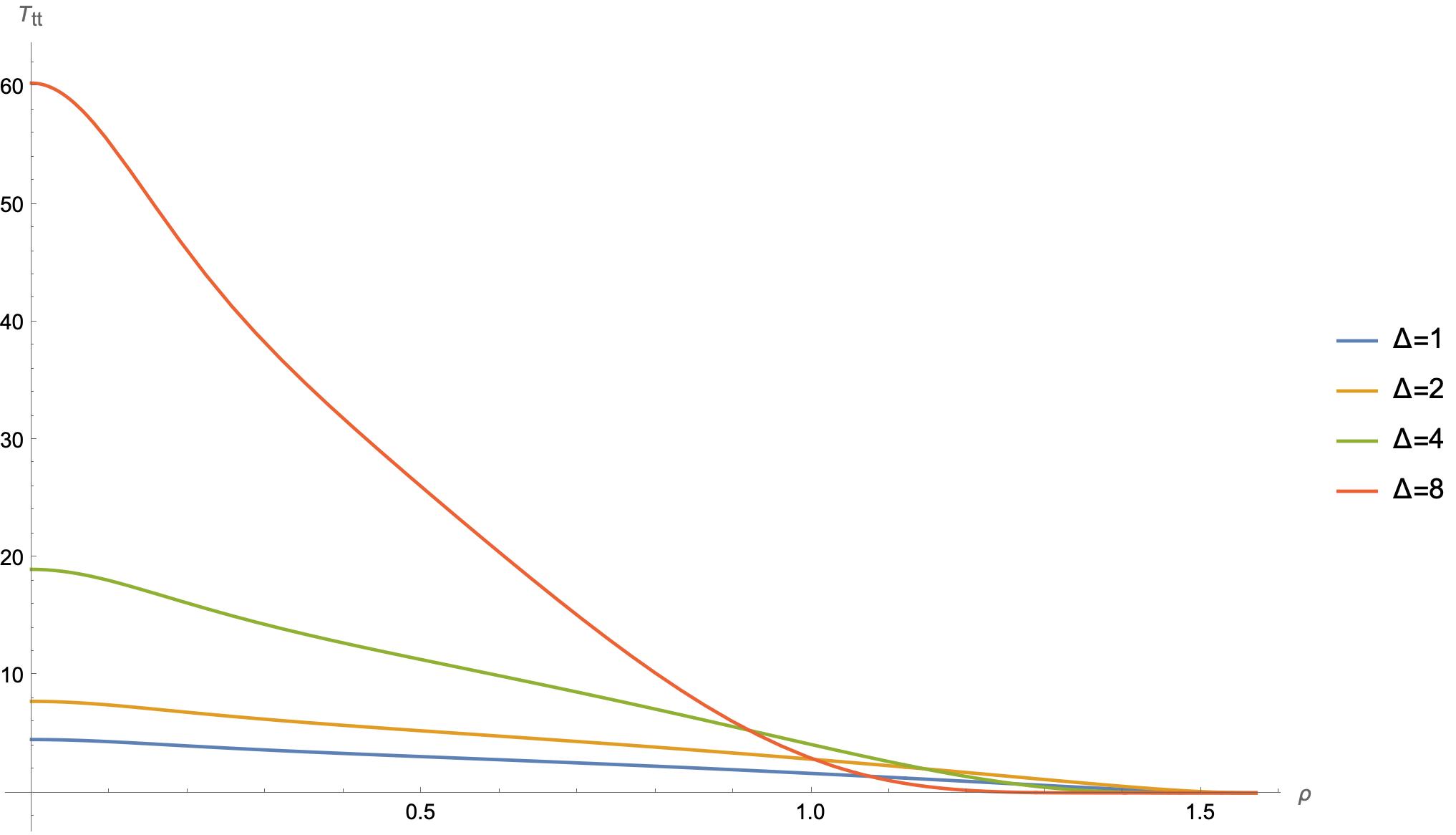}
    \caption{Stress tensor in the thermal ensemble for different values of $\beta$ and $\Delta$.}
    \label{fig:stress_tensor_1}
\end{figure}

\begin{figure}
 \centering
    \includegraphics[scale = 0.2]{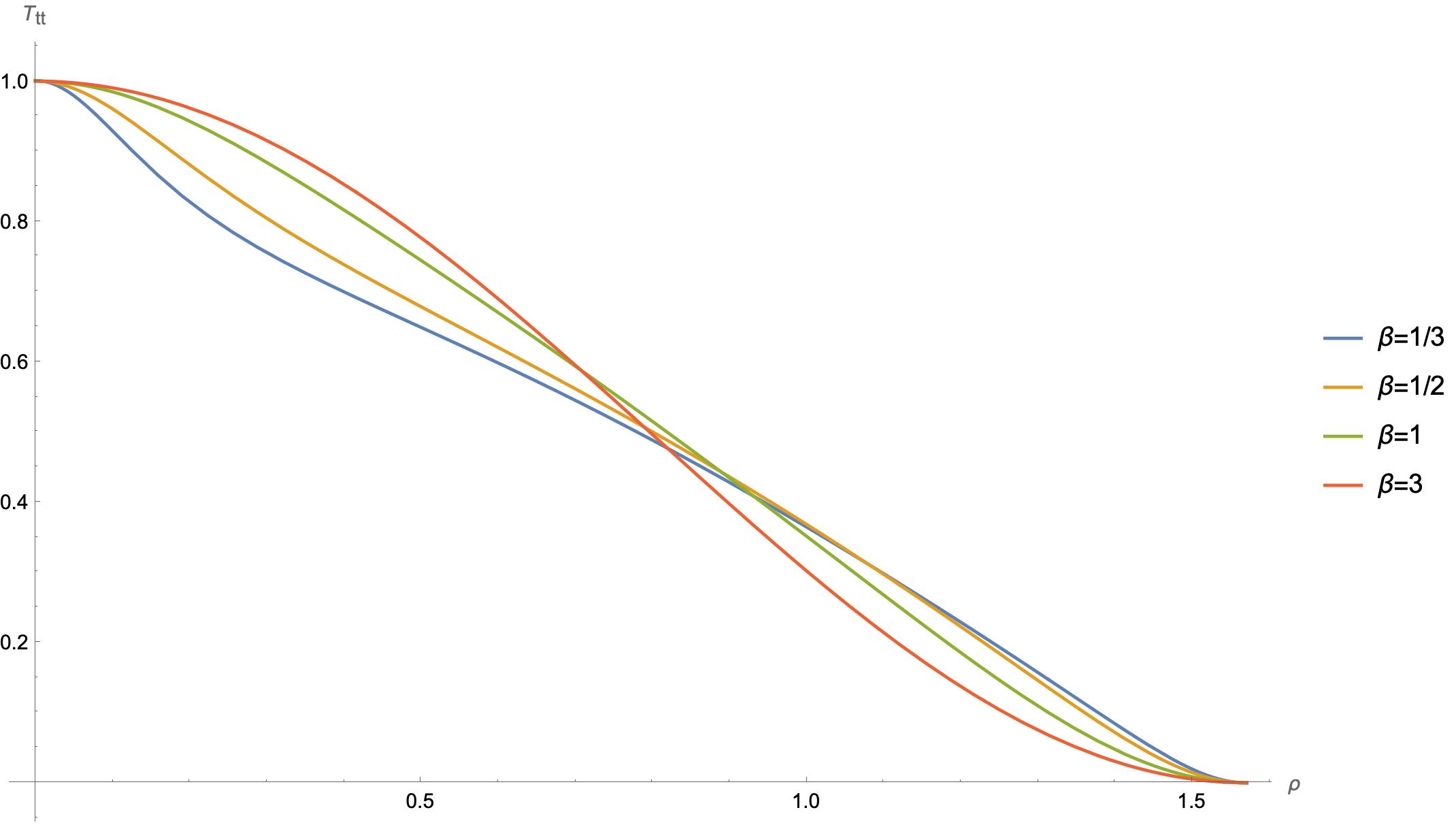} \hspace{0.5cm} \includegraphics[scale = 0.2]{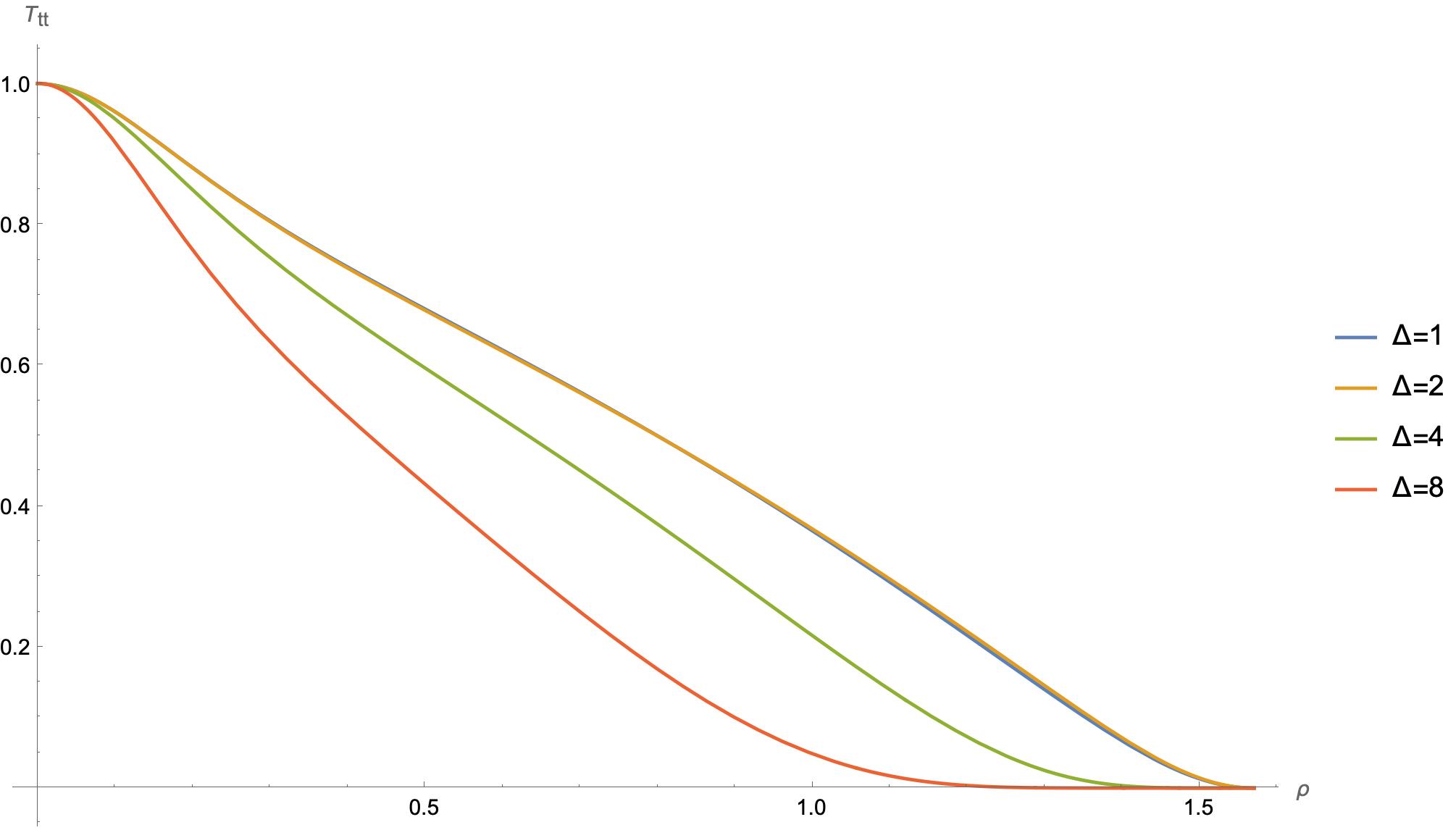}
    \caption{Normalised stress tensor in the thermal ensemble for different values of $\beta$ and $\Delta$. We normalise by the value of $\langle T_{tt} \rangle$ by its value at $\rho = 0$.}
    \label{fig:stress_tensor_2}
\end{figure}

We can also look at the profile of stress-tensor as we move from the bulk to the boundary. In figure \ref{fig:stress_tensor_1}, we plot $\langle T_{tt} \rangle_{\beta}$ for $\rho\in[0,\pi/2)$. Where by $\langle T_{tt} \rangle_{\beta}$ we mean the expectation value in the thermal ensemble at an inverse temperature $\beta$. From these plots, we clearly see that for large $\Delta$, the energy density increases around $\rho=0$ as we consider large values of $\Delta$ or temperatures. These limits also lead to a more localised energy densities. To compare the plots for different parameters, in figure \ref{fig:stress_tensor_2}, we plot a normalised stress where we divide the thermal expectation value $\langle T_{tt} \rangle_{\beta}$ by its value at the origin $\left \langle T_{tt} \rangle_{\beta} \right|_{\rho=0}$.

\section{More scalar wormholes and cosmologies}

\subsection{Sphere boundaries}
\label{sec:sphere}
Here, we describe wormhole solutions with spatially-dependent scalars where the spatial slices are two-spheres rather than flat space. In this the metric takes the form 
\begin{equation}
    ds^2 = dt^2 + a(\tau)^2 d\Omega_d^2
\end{equation}
The Einstein tensor for this metric is
\begin{equation}
    G_{\mu\nu} + \Lambda g_{\mu\nu} = 
        \left(
        \begin{array}{ccc}
         \frac{\dot a(t)^2-1}{a(t)^2}-1 & 0 & 0 \\
         0 & a(t) \left(\ddot a(t)-a(t)\right) & 0 \\
         0 & 0 & a(t) \sin ^2(\theta ) \left(\ddot a(t)-a(t)\right) \\
        \end{array}
        \right) 
\end{equation}
The complex scalar field solutions above section suggest a nice ansatz for the scalar field solution in the spherical case. In the flat case, the part $e^{ikx}$ are the basis modes to expand arbitrary function consistent with the symmetries of flat space. To write down solutions for the sphere, we then consider
\begin{equation}
    \Psi_{lm} = \sqrt{\pi}\, h(t) Y_{lm}(\theta,\psi) \,, \quad (l - \mbox{fixed}\,, m=-l \mbox{ to } l)
\end{equation} 
The stress tensor for these fields is 
\begin{equation}
    T_{\mu\nu} = 
        \left(
        \begin{array}{ccc}
         \frac{2l+1}{4} \left(\dot h(t)^2 - \frac{l(l+1) h(t)^2}{a(t)^2}\right) & 0 & 0 \\
         0 & -\frac{2l+1}{4} a(t)^2 \dot h(t)^2 & 0 \\
         0 & 0 & -\frac{2l+1}{4}  a(t)^2 \sin ^2(\theta ) \dot h(t)^2 \\
        \end{array}
        \right)
\end{equation}
The equation of motion are 
\begin{equation}
    \ddot h +  2\left(\frac{\dot a}{a}\right) \dot h - \frac{l(l+1)}{a^2}h = 0 \,, \quad \mbox{(Scalar EOM)}
\end{equation}
\begin{equation}
    \frac{\dot a^2}{a^2} = 1 + \frac{1}{a^2} + \frac{\kappa(2l+1)}{4} \left(\dot h^2 - \frac{l(l+1) h^2}{a^2} \right) \,, \quad \mbox{(Metric EOM)}
\end{equation}
\begin{equation}
    \frac{\ddot a}{a} = 1 - \frac{\kappa (2l + 1)}{4} \dot h^2
\end{equation}
These equations look similar to the case of flat space but with $k$ replaced by $l(l+1)$ and an additional term $1/a^2$, in the metric equation  due to the curvature of the sphere. One can perform a rescaling to make these equations look simpler.
\begin{equation}
    a \rightarrow \sqrt{l(l+1)} \, a \,\,, \quad  h \rightarrow 2h/\sqrt{\kappa (2l+1)}
\end{equation}
\begin{equation}
    \ddot h +  2\left(\frac{\dot a}{a}\right) \dot h - \frac{h}{a^2} = 0 
\end{equation}
\begin{equation}
    \frac{\dot a^2}{a^2} = 1 + \frac{1}{l(l+1) a^2} + \left(\dot h^2 - \frac{h^2}{a^2} \right) 
\end{equation}
\begin{equation}
    \frac{\ddot a}{a} = 1 - \dot h^2
\end{equation}
Such a rescaling would have removed all parameters $(k,\kappa)$ from the differential equations in the flat case. However, for the case of sphere this is no longer true. Due to the extra term (\textcolor{blue}{spatial curvature}) one still has the dependence on $l$. Note that in  the limit $l\rightarrow \infty$ (small wavelengths of the scalar fields) these reduce to the case of flat eoms.

Again, we look for solutions such that $a(t)$ and $h(t)$ are even functions in $t$ around $0$ with the initial conditions
\begin{equation}
    a(0) = a_0 \,,\quad \dot a(0) = 0 \,, \quad h(0) = \pm\sqrt{a_0^2 + \frac{1}{l(l+1)} } \,, \quad \dot h(0) = 0
\end{equation}

\subsection{WKB solution}
\label{sec:WKB}

In this section, we note that the equations (\ref{cosmo1}) and (\ref{cosmo2}) for cosmologies supported by scalar fields with a sinusoidal spatial profile have a WKB solution that provides a good approximation for short wavelengths.

If we define $s$ via
\begin{equation}
    {d \over ds} = a^3 {d \over dt}
\end{equation}
then the scalar equation becomes
\begin{equation}
{d h \over ds^2} = -p^2(s) h 
\end{equation}
where 
\begin{equation}
    p(s) = a(t)^2 \sqrt{k^2 + m^2 a^2(t)} \; .
\end{equation}
This takes the form of a (analytically continued) Schrodinger equation. When $k$ is large compared with the scale associated with the variation of $a(t)$, the solution should be well-approximated by the WKB solution
\begin{equation}
    h = {{ \cal C} \over \sqrt{p(s)}}e^{\pm i\int p(s) ds} \; .
\end{equation}
In terms of the original Euclidean time coordinate (choosing the even solution), we have
\begin{equation}
    h \approx {{\cal C} \over a(t) (k^2 + m^2 a^2(t))^{1 \over 4} }\cos \left(\int_0^t d t' \sqrt{k^2/a(t')^2+ m^2 } \right) \; .
\end{equation}
With this solution, we have approximately
\begin{equation}
     {1 \over 2}(\dot{h})^2 - {1 \over 2}\left(m^2 + {k^2 \over a^2}\right) h^2 \approx {1 \over 2} {{\cal C}^2 \over a^4}
\end{equation}
where we are keeping the $t$ derivatives from the $\cosh$ but dropping the rest. Plugging back into the Friedmann equation, we get simply
\begin{equation}
     \left({\dot{a} \over a}\right)^2 = -1 +{1 \over 2} {{\cal C}^2 \over a^4}
\end{equation}
Choosing $a=1$ at $t=0$ fixes ${\cal C}$, and we can simplify to 
\begin{equation}
     \left({\dot{a} \over a}\right)^2 = -1 + {1 \over a^4}
\end{equation}
which has solution
\begin{equation}
    a(t) = \sqrt{\cos(2t)} \; .
\end{equation}
This is the analytic continuation of the standard $\Lambda < 0$ plus radiation cosmology.

We have checked that this is a very good approximation to the numerical solution provided that $k \gg 1$.
On the other hand, this WKB solution is {\it not} a good approximation in the wormhole case.

\bibliographystyle{JHEP}
\bibliography{Hyperbolic}

\end{document}